\documentclass[prb,onecolumn,superscriptaddress,showpacs, nofootinbib]{revtex4-2}
\usepackage{graphicx}
\usepackage{float}
\usepackage{dcolumn}
\usepackage{float}
\usepackage{bm}
\usepackage{mathrsfs}
\usepackage{ulem}
\usepackage[per-mode = fraction]{siunitx}
\usepackage{hyperref}
\usepackage{xcolor}
\usepackage{soul}
\usepackage{amssymb}
\usepackage{amsthm}
\usepackage{mathtools}
\begin{document}

\title{Multiple-q current states in a multicomponent superconducting channel}
\author{Yuriy Yerin}
\affiliation{ Dipartimento di Fisica e Geologia, Universit\'a degli Studi di Perugia, Via Pascoli, 06123 Perugia, Italy}
\affiliation{CNR-SPIN, via del Fosso del Cavaliere, 100, 00133 Roma, Italy}
\author{Stefan-Ludwig Drechsler}
\affiliation{Institute for Theoretical Solid State Physics, Leibniz-Institut für Festkörper- und Werkstoffforschung IFW-Dresden, D-01169 Dresden, Helmholtzstraße 20}
\author{Mario Cuoco}
\affiliation{CNR-SPIN, c/o Universit\'a di Salerno, I-84084 Fisciano (SA), Italy}
\author{Caterina Petrillo}
\affiliation{Dipartimento di Fisica e Geologia, Universit\'a degli Studi di Perugia, Via Pascoli, 06123 Perugia, Italy}
\date{\today}

\begin{abstract}
It is well-established that multicomponent superconductors can host different nonstandard phenomena such as broken-time reversal symmetry (BTRS) states, exotic Fulde-Ferrell-Larkin-Ovchinnikov (FFLO) phases, the fractional Josephson effect as well as plenty of topological defects like phase solitons, domain walls and unusual vortex structures. We show that in the case of a two-component superconducting quasi-one-dimensional channel this catalogue can be extended by a novel inhomogeneous current state, which we have termed as a multiple-momenta state or, in short, a multiple-q state, characterized by the coexistence of two different interpenetrating Cooper pair condensates with different total momenta. Within the Ginzburg-Landau formalism for a dirty two-band superconductor with sizable impurity scattering treated in the Born-approximation we reveal that under certain conditions, the occurrence of multiple-q states can induce a cascade of transitions involving switching between them and the homogeneous BTRS (non-BTRS) states and vice versa leading this way to a complex interplay of homogeneous and inhomogeneous current states. We find that hallmarks of such a multiple-q state within a thin wire or channel can be a saw-like dependence of the depairing current and the existence of two distinct stable branches on it (a bistable current state). 
\end{abstract}
\maketitle

\section{Introduction}

The study of multicomponent superconductivity has become one of the major research  topics of condensed matter physics. The attention to this issue stems primarily from the fact that multicomponent superconductivity reveals a field with significantly rich physics and new interesting phenomena and unusual states not observed in conventional superconductors. The variety of multicomponent superconducting systems is represented for instance by
 strontium ruthenate \cite{Maeno, Kallin}, iron-based \cite{Tafti, Mazin, Klauss}, noncentrosymmetric \cite{Smidman} and heavy-fermions superconductors \cite{Izawa}. Loosely speaking, these materials can be considered as a kind  of stage theater where  the actors can play the role of various exotic states and phenomena. In this regard, it is evident that much effort and research is being made to discover and cast new promising “actors”, viz.\ new phenomena and states, unknown until now in multicomponent superconductors.

Among the effects that have already been discovered and those that are yet to be discovered, a special niche is occupied by the so-called phase coherent effects in multicomponent superconductors, connected with the emergence of nontrivial phase shifts between several distinctive order parameters. This can lead to the interesting phenomenon known as chiral superconductivity with $s_{\pm}+is_{++}$ pairing symmetry and as a consequence to state broken time-reversal symmetry (BTRS), when the phases of the multicomponent order parameter do exhibit frustration. 

The presence of non-zero phase shifts raises a reasonable question, namely how are these topics they manifested or could they become visible in the observables? At this stage, it has already been theoretically established that the occurrence of such phase difference topics should affect the Josephson effect with the appearance of  $\phi$ ($\phi_0$) and $\pi$ junctions and the corresponding current-phase relations \cite{Buzdin_phi, Ng, Yerin1, Yerin2, Yerin3, Yerin4, Guarcello1, Moor, Sasaki, Grigorishin1}, phase-sensitive structures like dc-SQUID with the unusual Fraunhofer diffraction patterns \cite{Yerin5}, the Little-Parks effect with the non-parabolic dependence of the critical temperature shift \cite{Yerin6, Askerzade1} and current states with anomalous characteristics of depairing curves \cite{Yerin7}. Moreover, under certain circumstances an applied magnetic flux can drive the phase shift, converting a state with chiral $s_{\pm}+is_{++}$ symmetry into a $s_{\pm}$ configuration, when the intercomponent phase difference is stable and equal to $\pi$, and vice versa \cite{Yerin8}. Such a controlled switching between current states of different symmetries (different phase shits) can produce an anomalous diamagnetic response inducting current density jumps and kinks in doubly-connected geometries \cite{Yerin9}. 

Besides, the intercomponent phase difference itself can arise due to topological excitations inherent solely in multicomponent superconductors and known as phase solitons of the sine-Gordon type \cite{Tanaka2002, Babaev2002, Yerin10, Vakaryuk, Lin2, Samokhin2} or double sine-Gordon type \cite{Yerin11}. These inhomogeneous current states have been confirmed experimentally in a series of experiments \cite{Bluhm, Tanaka1, Tanaka2, Tanaka3}. Another example of inhomogeneous current state is the Fulde-Ferrell-Larkin-Ovchinnikov (FFLO) state in a two-band superconductor, when due to the competition of two different modulation length scales, the FFLO phase is transformed into two phases separated by a first order phase transition: the so-called $Q_1$- and $Q_2$-FFLO phases at the higher and lower fields \cite{Machida1, Machida2}. 

This similarity obviously suggests the possibility for the existence of a current state in a multicomponent superconductor, in which different coexisting condensates will have different superconducting momenta $q_i$, where $i$ is the number of the component.  Strictly speaking, such a situation, when condensates can have different momenta is not new and can be achieved theoretically by means of the additional contribution from the Andreev-Bashkin effect, when the  intercomponent current-current coupling gives rise to a dissipationless drag (also known as  entrainment) between the two components within a mixture of two superfluids \cite{Bashkin, Fil, Nespolo} or a superconductor coupled to a superfluid \cite{Alford} or between the neutron and proton condensates in the core of a neutron star \cite{Babaev_star, Wood}.

In this paper, we intend to demonstrate that an inhomogeneous current state with different superconducting condensate momenta can arise in a superconductor even without taking into account the Andreev-Bashkin current-current coupling. In the framework of the Ginzburg-Landau phenomenological theory, it will be shown that a two-band superconducting quasi-one-dimensional channel with a weak interaction between the bands and with the inclusion of the interband scattering effect is sufficient for the onset of such a state. Along with this we find that its occurrence can start a cascade of transitions between it and homogeneous states with breaking of the time-reversal symmetry and its preservation. In the context of unconventional  superconductivity with $d$-wave symmetry the coexistence of strong impurity scattering and $q$-dependent inhomogeneities induced at high-magnetic fields manifested in the celebrated  Fulde-Ferrel-Larkin-Ovchinnikov (FFLO) phases has been demonstrated recently \cite{Agterberg2001,Vorontsov2008}. Here, we will show that a similar modulation of the order parameters is also possible at low or ambient magnetic fields at least in thin wires or channels but induced by an external current for dirty two-band superconductors with chiral $s_{\pm}$+i$s_{++}$-symmetry.

The outline of the paper is as follows. In Sec.\ \ref{Sec:Model} we describe the geometrical characteristics of a channel and introduce the Ginzburg-Landau  (GL) formalism generalized for the case of a two-component order parameter with the interband scattering effect included. In Sec.\ \ref{Sec:Diagram} we study the phase diagram of a two-component superconductor, where domains with a nontrivial phase difference as a function of the temperature and the strength of the interband scattering rate are shown. In this phase diagram we select reference points from each domain, which are the basis for the presentation of our results and subsequent conclusions. Following this, in Sec.\ \ref{sec:Transitions} we derive general expressions for the GL free energy and investigate its behavior  for selected reference points. The results of our calculations are discussed in Sec.\ \ref{sec:Discussion}. Finally, we present our conclusions in Sec.\ \ref{sec:Conclusion}.

\section{Model and formalism}
\label{Sec:Model}

\begin{figure}
\includegraphics[width=0.79\columnwidth]{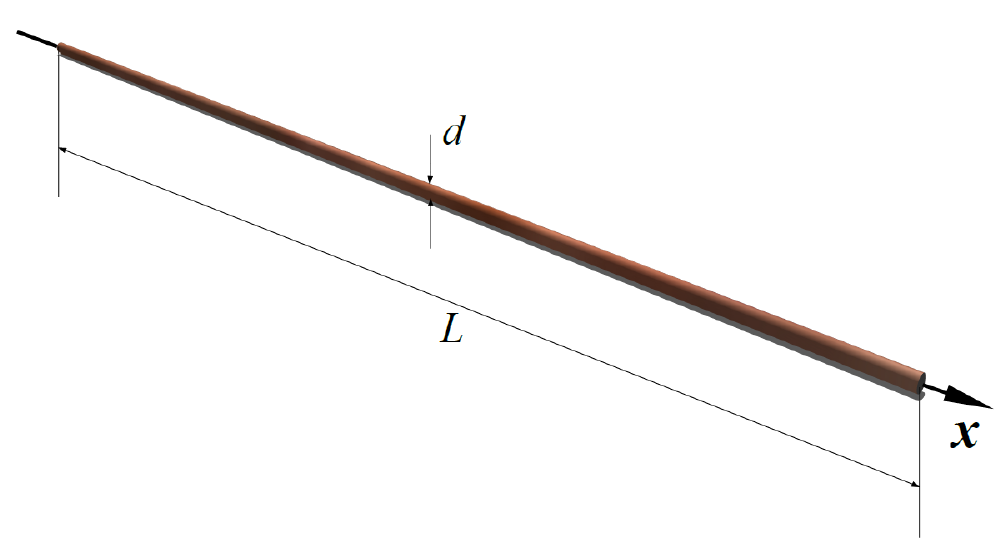}
\caption {The long thin wire with the thickness $d$ and the length $L$ is the proposed experimental system under consideration to reveal its multiple-$q$ character by measuring the current. Thereby it is assumed that the length of the wire or channel much exceeds the coherence length of the dirty two-band superconductor  $L \gg \xi(T)$  to guarantee the rare (usually {\it nonuniversal}!) equivalence of voltage and current driven responses (see Ref.\ \onlinecite{McCumber}).}
\label{wire}
\end{figure}

The subject of our consideration is given by the current states in a thin two-band superconducting wire with the diameter $d \ll \xi_{1, 2} (T), \lambda_{1, 2} (T)$, where $\xi_{1, 2}(T)$ and $\lambda_{1, 2}(T)$ are coherence lengths and London penetration depths for each non-interacting order parameter, respectively (Fig. \ref{wire}).

The research tool for the study of current states will be the the GL-theory for a dirty two-band superconductor. For this physical case, by means of the Usadel equations generalized for a two-band superconductor with interband scattering by impurities one can deduce the free energy $F$ \cite{Stanev, Corticelli} to the form
\begin{equation}
\label{GL_general}
F = {F_1} + {F_2} + {F_{12}} + \int{\frac{{{{\left( {{\text{rot }}{\mathbf{A}} - {\mathbf{H}}} \right)}^2}}}{{8\pi }}} {d^3}{\mathbf{r}},
\end{equation}
where $F_i$ are the partial contributions of the \textit{i}th band, $F_{12}$ is the component arising from the interband interaction which is also affected by the presence of interband impurity scattering. The last term describes the contribution of a magnetic field $\mathbf{H}$ and the vector-potential $\mathbf{A}$. The expressions for $F_i$ and $F_{12}$ have the form
\begin{widetext}
\begin{equation}
\label{GL_general1}
{F_1} = \int{\left[ {{a_{11}}{{\left| {{\Delta _1}} \right|}^2} + \frac{1}{2}{b_{11}}{{\left| {{\Delta _1}} \right|}^4} + \frac{1}{2}{k_{11}}{{\left| { - i\hbar \nabla  - \frac{{2e}}{c}{\mathbf{A}}} \right|}^2}{\Delta _1}} \right]} {d^{\mathbf{3}}}{\mathbf{r}},
\end{equation}
\begin{equation}
\label{GL_general2}
{F_2} = \int{\left[ {{a_{22}}{{\left| {{\Delta _2}} \right|}^2} + \frac{1}{2}{b_{22}}{{\left| {{\Delta _2}} \right|}^4} + \frac{1}{2}{k_{22}}{{\left| { - i\hbar \nabla  - \frac{{2e}}{c}{\mathbf{A}}} \right|}^2}{\Delta _2}} \right]} {d^{\mathbf{3}}}{\mathbf{r}},
\end{equation}
\begin{equation}
\label{GL_general3}
\begin{gathered}
  {F_{12}} = \int{\left[ {{b_{12}}{{\left| {{\Delta _1}} \right|}^2}{{\left| {{\Delta _2}} \right|}^2}} \right.}  + 2\left( {{a_{12}}\left| {{\Delta _1}} \right|\left| {{\Delta _2}} \right| + {c_{11}}{{\left| {{\Delta _1}} \right|}^3}\left| {{\Delta _2}} \right| + {c_{22}}\left| {{\Delta _1}} \right|{{\left| {{\Delta _2}} \right|}^3}} \right)\cos \phi  + {c_{12}}{\left| {{\Delta _1}} \right|^2}{\left| {{\Delta _2}} \right|^2}\cos 2\phi  \hfill \\
  \left. { + \frac{1}{2}{k_{12}}\left( {\left( { - i\hbar \nabla  - \frac{{2e}}{c}{\mathbf{A}}} \right){\Delta _1}\left( {i\hbar \nabla  - \frac{{2e}}{c}{\mathbf{A}}} \right)\Delta _2^* + \left( {i\hbar \nabla  - \frac{{2e}}{c}{\mathbf{A}}} \right)\Delta _1^*\left( { - i\hbar \nabla  - \frac{{2e}}{c}{\mathbf{A}}} \right){\Delta _2}} \right)} \right]{d^3}{\mathbf{r}}. \hfill \\ 
\end{gathered} 
\end{equation}
\end{widetext}
Here, ${\Delta _i} = \left| {{\Delta _i}} \right|\exp \left( {i{\chi _i}} \right)$ are complex order parameters. Also, we introduce the phase difference between the order parameters $\phi  = {\chi _2} - {\chi _1}$, which will play an important role for the determination of the ground state of a dirty two-band superconductor and for the description of the current states. 

The functional derivative $\frac{\partial F}{\partial \boldsymbol{A}(\boldsymbol{r})}$ yields the expression for the current $\boldsymbol{j}$:
\begin{equation}
\label{current_general}
\begin{array}{l}
{\bf{j}} =  - ie\hbar {k_{11}}\left( {\Delta _1^*\nabla {\Delta _1} - {\Delta _1}\nabla \Delta _1^*} \right) - ie\hbar {k_{22}}\left( {\Delta _2^*\nabla {\Delta _2} - {\Delta _2}\nabla \Delta _2^*} \right) - ie\hbar {k_{12}}\left( {\Delta _1^*\nabla {\Delta _2} - {\Delta _2}\nabla \Delta _1^* - {\Delta _1}\nabla \Delta _2^* + \Delta _2^*\nabla {\Delta _1}} \right)\\
 - \frac{{4{e^2}}}{c}\left( {{k_{11}}{{\left| {{\Delta _1}} \right|}^2} + {k_{22}}{{\left| {{\Delta _2}} \right|}^2} + {k_{12}}\left( {\Delta _1^*{\Delta _2} + \Delta _2^*{\Delta _1}} \right)} \right){\bf{A}}.
\end{array}
\end{equation}

The microscopic expressions for the coefficients of the GL free energy functional are given in the Appendix \ref{sec:A}. Noteworthy, the coefficients $b_{12}$, $c_{ij}$ and $k_{12}$ in Eq.\ (\ref{GL_general3}) are absent in the case of a clean two-band superconductor. Their emergence is the result of the contribution of the interband impurities, whose strength is characterized by the interband scattering rate $\Gamma$, being proportional to the impurity concentration.

The special geometry of the system under consideration allows us to reduce the analysis of  the current states to a one-dimensional problem and to neglect the self-magnetic field of the wire. In the absence of external magnetic fields the calibration ${\mathbf{A}}=0$ is applied.

From the physical point of view the derivatives of the order parameter phases ${\frac{{d{\chi _1}}}{{dx}}}$ and ${\frac{{d{\chi _2}}}{{dx}}}$ determine the superfluid momenta of Cooper pairs. For a conventional superconductor or the so-called Fulde-Ferrell-Larkin-Ovchinnikov (FFLO) superconductor the modulation of the order parameter is described by a {\it single} plane wave (FF state) or a simple $\cos $-term (LO state) as its real part in the simplest cases. Here, for the thin wire or channel,  the vector of the detrimental for superconductivity depairing current plays a similar role as the strong magnetic field in the FFLO states in the bulk: it causes modulations of the order parameters to minimize its detrimental influence. This common effect rests on the special equivalence of voltage and current driven responses in the present experimental situation \cite{McCumber}.

We should make an important remark about the present form of the GL free energy. Since our analysis is based on Eq.\ (\ref{GL_general}) this approach is applicable to systems for the so-called voltage-driven regime. For the current-driven regime the study of current states should be performed by means of the Gibbs free energy with the additional contribution of the current $I$, because the phase difference between the ends of the wire (i.e.\  parameter $q$) becomes a dependent variable and is determined by the depairing $I$ \cite{McCumber, Samokhin1}.

The GL-equations for the order parameter will be derived in the following sections for different states.

\subsection{The GL-formalism for the BTRS state}
The calculation of the functional derivatives $\partial F/\partial \phi = 0$, $\partial F/\partial |\Delta_1| = 0$ and $\partial G/\partial |\Delta_2| = 0$ leads to equations for $|\Delta_i|$ and allows us to obtain solutions for their phase difference $\phi$ 
\begin{equation}
\label{GL_eq1_BTRS}
\begin{array}{l}
\left( {{a_{11}} + \frac{\displaystyle {{k_{11}}{\hbar ^2}{q^2}}}{\displaystyle 2}} \right)\left| {{\Delta _1}} \right| + {b_{11}}{\left| {{\Delta _1}} \right|^3} + {b_{12}}\left| {{\Delta _1}} \right|{\left| {{\Delta _2}} \right|^2} + \left( {{a_{12}} + \frac{\displaystyle {{k_{12}}{\hbar ^2}{q^2}}}{2} + 3{c_{11}}{{\left| {{\Delta _1}} \right|}^2} + {c_{22}}{{\left| {{\Delta _2}} \right|}^2}} \right)\left| {{\Delta _2}} \right|\cos \phi \\
\end{array}
\end{equation}
\begin{equation}
\label{GL_eq2_BTRS}
\begin{array}{l}
\left( {{a_{22}} + \frac{\displaystyle {{k_{22}}{\hbar ^2}{q^2}}}{2}} \right)\left| {{\Delta _2}} \right| + {b_{22}}{\left| {{\Delta _2}} \right|^3} + {b_{12}}{\left| {{\Delta _1}} \right|^2}\left| {{\Delta _2}} \right| + \left( {{a_{12}} + \frac{\displaystyle {{k_{12}}{\hbar ^2}{q^2}}}{2} + {c_{11}}{{\left| {{\Delta _1}} \right|}^2} + 3{c_{22}}{{\left| {{\Delta _2}} \right|}^2}} \right)\left| {{\Delta _1}} \right|\cos \phi \\
 + {c_{12}}{\left| {{\Delta _1}} \right|^2}\left| {{\Delta _2}} \right|\cos 2\phi  = 0,
\end{array}
\end{equation}

\begin{equation}
\label{phi_homo_sol1}
\sin \phi  = 0 \Rightarrow \phi  = 0,\phi  = \pi,
\end{equation}
which corresponds to $s_{++}$ and $s_{\pm}$ symmetry, respectively. The most interesting case is the BTRS solution with an arbitrary $\phi$ and the accompanied chiral symmetry $s_{\pm}+is_{++}$ 
\begin{equation}
\label{phi_homo_sol2}
\cos \phi  =  - \frac{\displaystyle {{{k_{12}}{\hbar ^2}}}{{}{{q}^2} + 2\left( {{a_{12}} + {c_{11}}{{\left| {{\Delta _1}} \right|}^2} + {c_{22}}{{\left| {{\Delta _2}} \right|}^2}} \right)}}{\displaystyle {4{c_{12}}\left| {{\Delta _1}} \right|\left| {{\Delta _2}} \right|}},
\end{equation}
which gives rise to two solutions for the phase difference and consequently leads to a kind of frustration with a two-fold degenerate ground states and a spontaneously broken ${\mathbb{Z}_2}$ time-reversal symmetry.

For $q=0$ one can derive analytical solutions for the amplitudes of the superconducting order parameters. There are two solutions which read
\begin{widetext}
\begin{equation}
\label{OP1_sol1}
\left| {\Delta _1^{\left( 0 \right)}} \right| ^2 =  - \frac{\displaystyle {{a_{11}}{b_{22}}{c_{12}} - {a_{11}}c_{22}^2 + {a_{12}}{b_{12}}{c_{22}} - {a_{12}}{b_{22}}{c_{11}} - {a_{12}}{c_{12}}{c_{22}} - {a_{22}}{b_{12}}{c_{12}} + {a_{22}}{c_{11}}{c_{22}} + {a_{22}}c_{12}^2}}{\displaystyle {{b_{11}}{b_{22}}{c_{12}} - {b_{11}}c_{22}^2 - b_{12}^2{c_{12}} + 2{b_{12}}{c_{11}}{c_{22}} + 2{b_{12}}c_{12}^2 - {b_{22}}c_{11}^2 - 2{c_{11}}{c_{12}}{c_{22}} - c_{12}^3}},
\end{equation}
\begin{equation}
\label{OP2_sol1}
\left| {\Delta _2^{\left( 0 \right)}} \right|^2 = \frac{\displaystyle {{a_{11}}{b_{12}}{c_{12}} - {a_{11}}{c_{11}}{c_{22}} - {a_{11}}c_{12}^2 + {a_{12}}{b_{11}}{c_{22}} - {a_{12}}{b_{12}}{c_{11}} + {a_{12}}{c_{11}}{c_{12}} - {a_{22}}{b_{11}}{c_{12}} + {a_{22}}c_{11}^2}}{\displaystyle {{b_{11}}{b_{22}}{c_{12}} - {b_{11}}c_{22}^2 - b_{12}^2{c_{12}} + 2{b_{12}}{c_{11}}{c_{22}} + 2{b_{12}}c_{12}^2 - {b_{22}}c_{11}^2 - 2{c_{11}}{c_{12}}{c_{22}} - c_{12}^3}},
\end{equation}
\end{widetext}
while for $q \ne 0$
\begin{equation}
\label{OP1_sol1q}
\begin{array}{l}
{\left| {{\Delta _1}} \right|^2}\left( q \right) = {\left| {\Delta _1^{\left( 0 \right)}} \right|^2}\\
 - \frac{1}{2}\frac{\displaystyle { - {b_{12}}{\mkern 1mu} {c_{12}}{\mkern 1mu} {k_{22}} + {b_{12}}{\mkern 1mu} {c_{22}}{\mkern 1mu} {k_{12}} - {b_{22}}{\mkern 1mu} {c_{11}}{\mkern 1mu} {k_{12}} + {b_{22}}{\mkern 1mu} {c_{12}}{\mkern 1mu} {k_{11}} + {c_{11}}{\mkern 1mu} {c_{22}}{\mkern 1mu} {k_{22}} + c_{12}^2{k_{22}} - {c_{12}}{\mkern 1mu} {c_{22}}{\mkern 1mu} {k_{12}} - c_{22}^2{k_{11}}}}{\displaystyle {{b_{11}}{b_{22}}{c_{12}} - {b_{11}}c_{22}^2 - b_{12}^2{c_{12}} + 2{b_{12}}{c_{11}}{c_{22}} + 2{b_{12}}c_{12}^2 - {b_{22}}c_{11}^2 - 2{c_{11}}{c_{12}}{c_{22}} - c_{12}^3}}{q^2},
\end{array}
\end{equation}
\begin{equation}
\label{OP2_sol2q}
\begin{array}{l}
{\left| {{\Delta _2}} \right|^2}\left( q \right) = {\left| {\Delta _2^{\left( 0 \right)}} \right|^2}\\
 - \frac{1}{2}\frac{\displaystyle { - {b_{11}}{\mkern 1mu} {c_{12}}{\mkern 1mu} {k_{22}} + {b_{11}}{\mkern 1mu} {c_{22}}{\mkern 1mu} {k_{12}} - {b_{12}}{\mkern 1mu} {c_{11}}{\mkern 1mu} {k_{12}} + {b_{12}}{\mkern 1mu} {c_{12}}{\mkern 1mu} {k_{11}} + c_{11}^2{k_{22}} + {c_{11}}{\mkern 1mu} {c_{12}}{\mkern 1mu} {k_{12}} - {c_{11}}{\mkern 1mu} {c_{22}}{\mkern 1mu} {k_{11}} - c_{12}^2{k_{11}}}}{\displaystyle {{b_{11}}{b_{22}}{c_{12}} - {b_{11}}c_{22}^2 - b_{12}^2{c_{12}} + 2{b_{12}}{c_{11}}{c_{22}} + 2{b_{12}}c_{12}^2 - {b_{22}}c_{11}^2 - 2{c_{11}}{c_{12}}{c_{22}} - c_{12}^3}}{q^2}.
\end{array}
\end{equation}

The subsequent substitution of the expression for the phase difference in the BTRS state given by Eq.\ (\ref{phi_homo_sol2}) into Eq.\ (\ref{Gibbs_final_BTRS}) yields a fourth-order polynomial of $q$
\begin{widetext}
\begin{equation}
\label{Gibbs_final_BTRS}
\begin{gathered}
  \frac{F}{{{L}}} = {F_0} - \frac{1}{{{c_{12}}}}\left[ {\frac{{{k_{12}^2\hbar ^4{q}^4}}}{{8{}}}{}} \right.
   + \left( {\left( {{c_{11}}{k_{12}} - {c_{12}}{k_{11}}} \right){{\left| {{\Delta _1}} \right|}^2} + \left( {{c_{22}}{k_{12}} - {c_{12}}{k_{22}}} \right){{\left| {{\Delta _2}} \right|}^2} + {a_{12}}{k_{12}}} \right)\frac{{{\hbar ^2q^2}}}{{2{}}}{} \hfill \\
  \left. { + \left( {\frac{1}{2}{a_{12}} + {c_{11}}{{\left| {{\Delta _1}} \right|}^2} + {c_{22}}{{\left| {{\Delta _2}} \right|}^2}} \right){a_{12}} + \frac{1}{2}{{\left( {{c_{11}}{{\left| {{\Delta _1}} \right|}^2} + {c_{22}}{{\left| {{\Delta _2}} \right|}^2}} \right)}^2} + c_{12}^2{{\left| {{\Delta _1}} \right|}^2}{{\left| {{\Delta _2}} \right|}^2}} \right]. \hfill \\ 
\end{gathered} 
\end{equation}
\end{widetext}

Eqs.\ (\ref{GL_eq1_BTRS}, \ref{GL_eq2_BTRS}) and (\ref{phi_homo_sol2}) must be supplemented an expression for the total current
\begin{equation}
\label{current_BTRS}
\frac{I}{L}=2e\hbar {k_{11}}{\left| {{\Delta _1}} \right|^2}q + 2e\hbar {k_{22}}{\left| {{\Delta _2}} \right|^2}q - e\hbar \frac{{{k_{12}}}}{{{c_{12}}}}q\left( {{k_{12}}{\hbar ^2}{q^2} + 2\left( {{a_{12}} + {c_{11}}{{\left| {{\Delta _1}} \right|}^2} + {c_{22}}{{\left| {{\Delta _2}} \right|}^2}} \right)} \right).
\end{equation}

\subsection{The GL-formalism for the homogeneous state}
For the homogeneous case we have
\begin{equation}
\label{GL_homo}
\begin{array}{l}
\frac{F}{L} = {F_0} + \left( {\frac{1}{2}{k_{11}}{{\left| {{\Delta _1}} \right|}^2} + \frac{1}{2}{k_{22}}{{\left| {{\Delta _2}} \right|}^2} + {k_{12}}\left| {{\Delta _1}} \right|\left| {{\Delta _2}} \right|\cos \phi } \right){\hbar ^2}{q^2}\\
 + 2\left( {{a_{12}}\left| {{\Delta _1}} \right|\left| {{\Delta _2}} \right| + {c_{11}}{{\left| {{\Delta _1}} \right|}^3}\left| {{\Delta _2}} \right| + {c_{22}}\left| {{\Delta _1}} \right|{{\left| {{\Delta _2}} \right|}^3}} \right)\cos \phi  + {c_{12}}{\left| {{\Delta _1}} \right|^2}{\left| {{\Delta _2}} \right|^2}\cos 2\phi.
\end{array}
\end{equation}

Correspondingly , the GL equations for the order parameters have the form
\begin{equation}
\label{GLeq1_homo}
\begin{array}{l}
\left( {{a_{11}} + \frac{\displaystyle {{k_{11}}{\hbar ^2}{q^2}}}{\displaystyle2}} \right)\left| {{\Delta _1}} \right| + {b_{11}}{\left| {{\Delta _1}} \right|^3} + {b_{12}}\left| {{\Delta _1}} \right|{\left| {{\Delta _2}} \right|^2} + \left( {{a_{12}} + \frac{\displaystyle{{k_{12}}{\hbar ^2}{q^2}}}{\displaystyle2} + 3{c_{11}}{{\left| {{\Delta _1}} \right|}^2} + {c_{22}}{{\left| {{\Delta _2}} \right|}^2}} \right)\left| {{\Delta _2}} \right|\cos \phi \\
 + {c_{12}}\left| {{\Delta _1}} \right|{\left| {{\Delta _2}} \right|^2}\cos 2\phi  = 0,
\end{array}
\end{equation}
\begin{equation}
\label{GLeq2_homo}
\begin{array}{l}
\left( {{a_{22}} + \frac{\displaystyle{{k_{22}}{\hbar ^2}{q^2}}}{\displaystyle2}} \right)\left| {{\Delta _2}} \right| + {b_{22}}{\left| {{\Delta _2}} \right|^3} + {b_{12}}{\left| {{\Delta _1}} \right|^2}\left| {{\Delta _2}} \right| + \left( {{a_{12}} + \frac{\displaystyle{{k_{12}}{\hbar ^2}{q^2}}}{\displaystyle2} + {c_{11}}{{\left| {{\Delta _1}} \right|}^2} + 3{c_{22}}{{\left| {{\Delta _2}} \right|}^2}} \right)\left| {{\Delta _1}} \right|\cos \phi \\
 + {c_{12}}{\left| {{\Delta _1}} \right|^2}\left| {{\Delta _2}} \right|\cos 2\phi  = 0,
\end{array}
\end{equation}
while for the total current
\begin{equation}
\label{current_homo}
\frac{I}{L} = 2e\hbar {k_{11}}{\left| {{\Delta _1}} \right|^2}q + 2e\hbar {k_{22}}{\left| {{\Delta _2}} \right|^2}q + 4e\hbar {k_{12}}\left| {{\Delta _1}} \right|\left| {{\Delta _2}} \right|q.
\end{equation}

\subsection{The GL-formalism for the multiple-$q$  state}

Strictly speaking, there are no convincing arguments against the assumption that the superconducting momenta of both condensates in a two-component superconductor can have different values, rather than one, as  introduced in the previous section for the homogeneous state. This implies that we can represent the order parameters as plane waves with different $q_1$ and $q_2$ wave vectors
\begin{equation}
\label{q1-q2_exp}
{\Delta _i} = \left| {{\Delta _i}} \right|\exp \left( {i{q_i}x} \right) \ .
\end{equation}

A similar approach with the introduction of two competing wave-vectors has been used for the study of the phase diagram of Pauli-limiting two-band superconductors  \cite{Machida1, Machida2}. There the emergence of the exotic FFLO state was predicted. The latter is divided into two states by a first-order transition: the $Q_1$- and $Q_2$-FFLO states at the higher and the lower magnetic field, respectively. Based on this similarity we term the inhomogeneous current state under consideration  "multiple-momenta state" that we abbreviate for convenience as a multiple-$q$ state. The term "multiple" was deliberately chosen because the state with two momenta can be generalized to the case of a superconductor in which the order parameter has more than two components. 

The present analytical consideration is rests for the lack of space and simplicity, so far, on the ansatz given by the Eq.\ (\ref{q1-q2_exp}), and we have not yet considered also for the same reason another possible {\it intrinsic},  closely related co-sinusoidally modulation of the order parameter like in the LO phase.  Moreover, we have also not addressed the interplay with other, {\it external} to our proposed mechanism, modulations such as pair density waves (PDW) [see for instance the comprehensive review by Agterberg {\it et al.} \cite{PDW}], which consideration itself represents a separate problem for future research. Anyhow, for all these interesting cases the problem of depairing currents in the corresponding one-band cases should be addressed first, which however has not yet been done so far for the best of our knowledge.

At a first glance this inhomogeneous multiple-$q$ state is reminiscent to phase soliton states in a multi-component superconductor \cite{Tanaka2002, Babaev2002, Yerin10, Vakaryuk, Lin2, Samokhin2}. Indeed, for the case of 
 thin-walled two-band superconducting cylinders with the radius $R$, the phase soliton is described by the sine-Gordon equation with the solution in terms of Jacobi elliptic functions \cite{Yerin10}
\begin{equation}
\label{cylinder_soliton}
\phi_{n}\left(  \varphi\right)  =\frac{\left(  1+\text{sgn}~a_{12}\right)}{2}+2\text{am}~\left(  \frac{nK\left(  k_{n}\right)  }{\pi}\left(\varphi-\varphi_{n0}\right)  ,k_{n}\right),
\end{equation}
where $am(u)$ denotes the elliptic amplitude, $K\left(  k\right)$ is the complete elliptic integral of the first kind, $\varphi$ is the polar coordinate, $\varphi_{n0}$ are arbitrary constants, and the $k_{n}$ ($n=\pm1,\pm2,\ldots$) satisfy the equations
\begin{equation}
\label{k_n}
\left\vert n\right\vert k_{n}K\left(  k_{n}\right)  =\frac{\pi R}{l}.
\end{equation}
Here the parameter $l$ is defined by internal properties of a two-band superconductor and is the inverse proportional to the interband interaction coefficient $a_{12}$, $l \simeq 1/\sqrt {\left| {{a_{12}}} \right|}$.
Based on Eq.\ (\ref{cylinder_soliton}) one can extract a particular example of the \textit{non}-soliton topological solution corresponding to the physical case of a very weak interband coupling. This solution can be obtained by expansions in series when $k_{n}\rightarrow0$ ($\frac{R}{l}\ll1$):
\begin{equation}
\label{weak_sol}
\phi_{n}\left(  \varphi\right)  \approx\frac{\left(  1+\text{sgn~}a_{12}\right)  \pi}{2}+n\left(  \varphi-\varphi_{0}\right),
\end{equation}
which for $a_{12}<0$ yields the dependence 
\begin{equation}
\label{weak_sol_new}
\phi_{n}\left(  \varphi\right)  \approx n\left(  \varphi-\varphi_{0}\right),
\end{equation}
that is similar to the introduced above phase difference $\phi(x)=(q_2-q_1)x$ (see Eq.\ (\ref{q1-q2_exp})) where the discrete number $n$ plays formally the role of a continuous variable $q_2-q_1$ and the shifted polar coordinate $ \varphi-\varphi_{0}$ is replaced for the Cartesian coordinate $x$ according to the geometry of the wire system under consideration (see Fig. \ref{wire}).

The substitution of Eq.\ (\ref{q1-q2_exp}) to Eq.\ (\ref{GL_general}) and subsequent integration over $x$ gives the GL free energy:
\begin{widetext}
\begin{equation}
\label{GL_q1-q2}
\begin{array}{c}
\frac{F}{L} = {F_0} + \frac{1}{2}{k_{11}}{\hbar ^2}{\left| {{\Delta _1}} \right|^2}q_1^2 + \frac{1}{2}{k_{22}}{\hbar ^2}{\left| {{\Delta _2}} \right|^2}q_2^2 + {k_{12}}{\hbar ^2}\left| {{\Delta _1}} \right|\left| {{\Delta _2}} \right|{q_1}{q_2}\frac{\displaystyle{\sin \left( {\left( {{q_1} - {q_2}} \right)L} \right)}}{\displaystyle{\left( {{q_1} - {q_2}} \right)L}}\\
 + 2\left( {{a_{12}}\left| {{\Delta _1}} \right|\left| {{\Delta _2}} \right| + {c_{11}}{{\left| {{\Delta _1}} \right|}^3}\left| {{\Delta _2}} \right| + {c_{22}}\left| {{\Delta _1}} \right|{{\left| {{\Delta _2}} \right|}^3}} \right)\frac{\displaystyle{\sin \left( {\left( {{q_1} - {q_2}} \right)L} \right)}}{\displaystyle{\left( {{q_1} - {q_2}} \right)L}} + \frac{1}{2}{c_{12}}{\left| {{\Delta _1}} \right|^2}{\left| {{\Delta _2}} \right|^2}\frac{\displaystyle{\sin \left( {2\left( {{q_1} - {q_2}} \right)L} \right)}}{\displaystyle{\left( {{q_1} - {q_2}} \right)L}},
\end{array}
\end{equation}
where 
\begin{equation}
	\label{F0_q1-q2}
	\begin{array}{c}
		{F_0} = {a_{11}}{\left| {{\Delta _1}} \right|^2} + {a_{22}}{\left| {{\Delta _2}} \right|^2} + \frac{1}{2}{b_{11}}{\left| {{\Delta _1}} \right|^4}
		+ \frac{1}{2}{b_{22}}{\left| {{\Delta _2}} \right|^4} + {b_{12}}{\left| {{\Delta _1}} \right|^2}{\left| {{\Delta _2}} \right|^2}.
	\end{array}
\end{equation}
\end{widetext}

After that we can perform the variation procedure and obtain the GL-equations
\begin{widetext}
\begin{equation}
\label{GLeq1_q1-q2}
\begin{array}{l}
\left( {{a_{11}} + \frac{\displaystyle {{k_{11}}{\hbar ^2}q_1^2}}{\displaystyle2}} \right)\left| {{\Delta _1}} \right| + {b_{11}}{\left| {{\Delta _1}} \right|^3} + {b_{12}}\left| {{\Delta _1}} \right|{\left| {{\Delta _2}} \right|^2} + \left( {{a_{12}} + \frac{\displaystyle{{k_{12}}{\hbar ^2}{q_1}{q_2}}}{\displaystyle2} + 3{c_{11}}{{\left| {{\Delta _1}} \right|}^2} + {c_{22}}{{\left| {{\Delta _2}} \right|}^2}} \right)\left| {{\Delta _2}} \right|\frac{\displaystyle{\sin \left( {\left( {{q_1} - {q_2}} \right)L} \right)}}{\displaystyle{\left( {{q_1} - {q_2}} \right)L}}\\
 + \frac{1}{2}{c_{12}}\left| {{\Delta _1}} \right|{\left| {{\Delta _2}} \right|^2}\ \frac{\displaystyle{\sin \left( {2\left( {{q_1} - {q_2}} \right)L} \right)}}{\displaystyle{\left( {{q_1} - {q_2}} \right)L}} = 0,
\end{array}
\end{equation}
\begin{equation}
\label{GLeq2_q1-q2}
\begin{array}{l}
\left( {{a_{22}} + \frac{\displaystyle {{k_{22}}{\hbar ^2}q_2^2}}{\displaystyle 2}} \right)\left| {{\Delta _2}} \right| + {b_{22}}{\left| {{\Delta _2}} \right|^3} + {b_{12}}{\left| {{\Delta _1}} \right|^2}\left| {{\Delta _2}} \right| + \left( {{a_{12}} + \frac{\displaystyle {{k_{12}}{\hbar ^2}{q_1}{q_2}}}{\displaystyle 2} + {c_{11}}{{\left| {{\Delta _1}} \right|}^2} + 3{c_{22}}{{\left| {{\Delta _2}} \right|}^2}} \right)\left| {{\Delta _1}} \right|\frac{\displaystyle {\sin \left( {\left( {{q_1} - {q_2}} \right)L} \right)}}{\displaystyle {\left( {{q_1} - {q_2}} \right)L}}\\
 + \frac{1}{2}{c_{12}}{\left| {{\Delta _1}} \right|^2}\left| {{\Delta _2}} \right|\frac{\displaystyle {\sin \left( {2\left( {{q_1} - {q_2}} \right)L} \right)}}{\displaystyle {\left( {{q_1} - {q_2}} \right)L}} = 0.
\end{array}
\end{equation}
\end{widetext}

In case of a very long channel, when  $L \to \infty$ is obeyed, one can ignore the terms with a sine function and find an approximate analytical solution of Eqs.\ (\ref{GLeq1_q1-q2}) and (\ref{GLeq2_q1-q2})
\begin{equation}
\label{GLeq1_q1-q2_sol}
{\left| {{\Delta _1}} \right|^2}\left( {{q_1}} \right) = \frac{{{a_{22}}{\mkern 1mu} {b_{12}} - {a_{11}}{\mkern 1mu} {b_{22}} - \left( {{b_{22}}{\mkern 1mu} {k_{11}} - {b_{12}}{\mkern 1mu} {k_{22}}} \right){\mkern 1mu} q_1^2}}{{{b_{11{\mkern 1mu} }}{b_{22}} - b_{12}^2}},
\end{equation}
\begin{equation}
\label{GLeq2_q1-q2_sol}
{\left| {{\Delta _2}} \right|^2}\left( {{q_2}} \right) = \frac{\displaystyle {{a_{11}}{\mkern 1mu} {b_{12}} - {a_{22}}{\mkern 1mu} {b_{11}} - \left( {{b_{11}}{\mkern 1mu} {k_{22}} - {b_{12}}{\mkern 1mu} {k_{11}}} \right){\mkern 1mu} q_2^2}}{\displaystyle {{b_{11{\mkern 1mu} }}{b_{22}} - b_{12}^2}}.
\end{equation}

For the characterization of the multiple-$q$ state it is necessary to provide the expression for the total current
\begin{equation}
\label{current_q1q2}
\frac{I}{L} = 2e\hbar {k_{11}}{\left| {{\Delta _1}} \right|^2}{q_1} + 2e\hbar {k_{22}}{\left| {{\Delta _2}} \right|^2}{q_2} + 2e\hbar {k_{12}}\left| {{\Delta _1}} \right|\left| {{\Delta _2}} \right|\left( {{q_1} + {q_2}} \right)\frac{\displaystyle {\sin \left( {{q_2} - {q_1}} \right)L}}{\displaystyle {\left( {{q_2} - {q_1}} \right)L}}.
\end{equation}

\section{Reference points on the phase diagram} 
\label{Sec:Diagram}
\begin{figure}
\includegraphics[width=0.99\columnwidth]{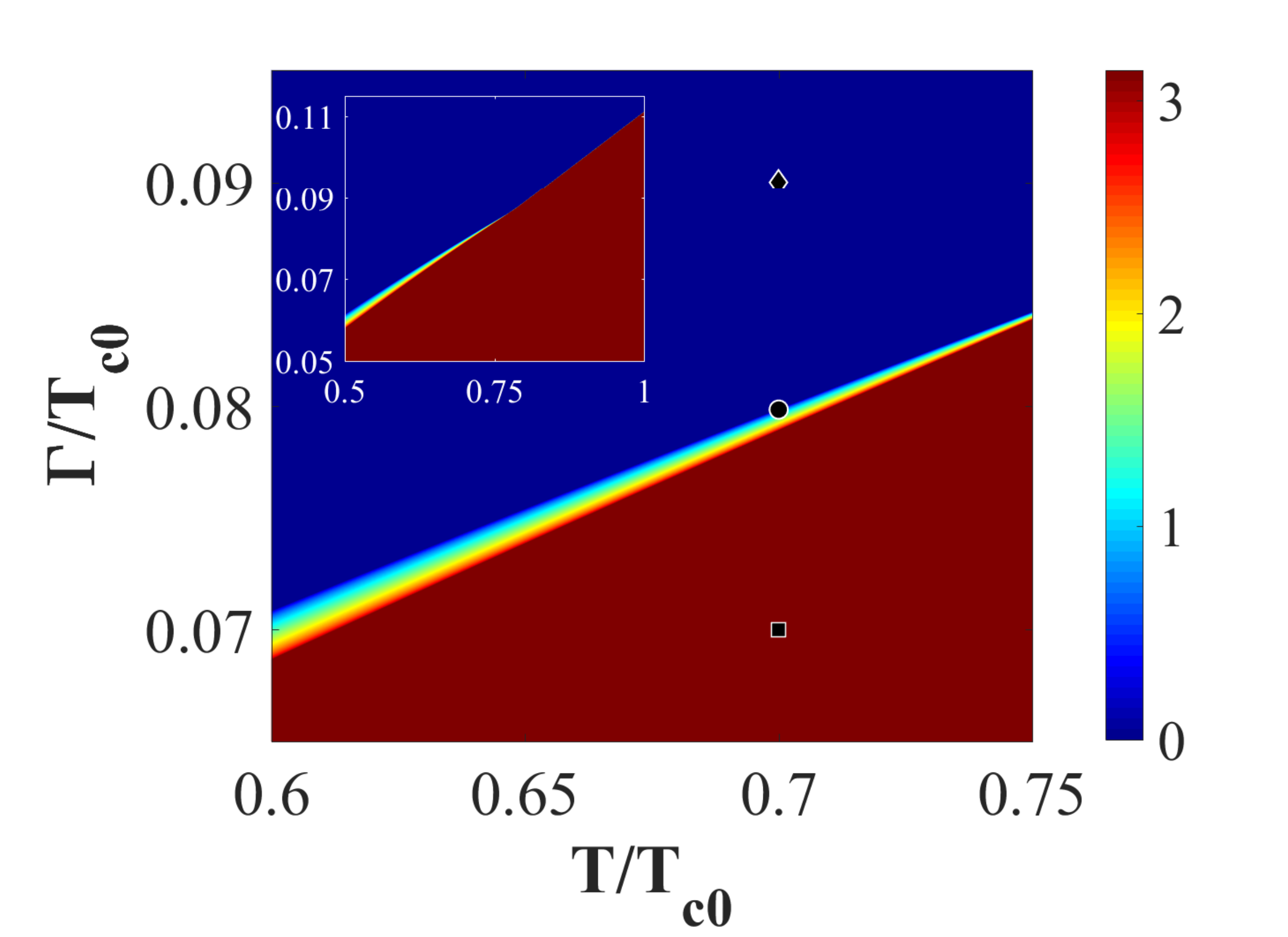}
\caption {The phase diagram for the phase difference $\phi$ (in radian units) as a function of the interband scattering rate $\Gamma$ and the temperature $T$ with the set of intra- and interband constants $\lambda_{11}=0.35$, $\lambda_{22}=0.347$,  $\lambda_{12}=\lambda_{21}=-0.01$. The narrow colorful domain represents the BTRS state with $s_{\pm}+is_{++}$ symmetry; blue and red domains stand for the non-BTRS state with $s_{\pm}$ and $s_{++}$ symmetry, respectively. The black filled square ($\Gamma/T_{c0}=0.07$), circle ($\Gamma/T_{c0}=0.07982$) and diamond ($\Gamma/T_{c0}=0.09$)  illustrate the reference points for the consideration of transitions between current states at $T/T_{c0}=0.7$. For the sake of clarity the inset shows a more extended view up to higher temperatures for the narrowing of the BTRS domain. $T_{c0}$ denotes the critical temperature of the reference parent clean system.}
\label{Phase_diag}
\end{figure}

 Eqs.\ (\ref{phi_homo_sol1}) and (\ref{phi_homo_sol2}) for $\phi$ together with the solutions for the order parameter from Eqs.\ (\ref{OP1_sol1}) and (\ref{OP2_sol1}) for the BTRS state allow to determine the phase difference as a function of the temperature and the interband scattering rate $\Gamma$ in the equilibrium phase when $q=0$ and $q_1=q_2=0$. Exploiting the microscopic expressions for the coefficients of GL free energy provided in the Appendix \ref{sec:A}, one can show the phase diagram with BTRS and non-BTRS domains. Figure \ref{Phase_diag} focuses on the phase diagram for a dirty two-band superconductor with the intraband $\lambda_{11}=0.35$, $\lambda_{22}=0.347$ and weak repulsive interband interaction constants $\lambda_{12}=\lambda_{21}=-0.01$, where the small cone-like colorful part illustrates a BTRS state with $\phi \ne 0$, while the large red and blue regions  for a non-BTRS state with $\phi =\pi$ and $\phi = 0$, respectively. 

To demonstrate the variety of current states and phase transitions between them in a superconducting quasi-one-dimensional wire, we choose three reference points on this phase diagram corresponding to different symmetries of the order parameter for the temperature $T/T_{c0}=0.7$. These starting points are marked on the phase diagram by the filled black square, circle and diamond (see Fig. \ref{Phase_diag}) and reflect three different types of symmetry of the dirty two-band superconductor. For $\Gamma=0.07T_{c0}$ (the filled black square) we have $s_{\pm}$ pairing symmetry  and a non-BTRS state ($\phi = \pi$). The point $\Gamma=0.07982T_{c0}$ (the filled black circle) is located on the upper edge of the BTRS state with $s_{\pm}+is_{++}$ chiral symmetry and $\phi \approx 2\pi/23$. Finally, for $\Gamma=0.09T_{c0}$ (the filled black diamond) $s_{++}$ symmetry and $\phi = 0$ is realized again with a non-BTRS state.

It is important to note that the our choice of the temperature $T/T_{c0}=0.7$ as well as the lower bound for the allowed temperature range specified in the phase diagram shown in Fig. \ref{Phase_diag} may be restricted  by the range of applicability of the GL theory for a dirty two-band superconductor. As a result, the microscopic theory for the description of the phase diagram should be applied \cite{Babaev_PD}. However, here we consider temperatures, which 
are sufficiently close to the $T_c$-values for the above selected values of the interband scattering rate (see Appendix \ref{sec:B} and Figure \ref{Tc_vs_Gamma} therein). Therefore, we suggest that our model calculations obey the 
validity of the phenomenological GL-approach.

\section{Phase transitions}
 \label{sec:Transitions}
\begin{figure}
\includegraphics[width=0.32\columnwidth]{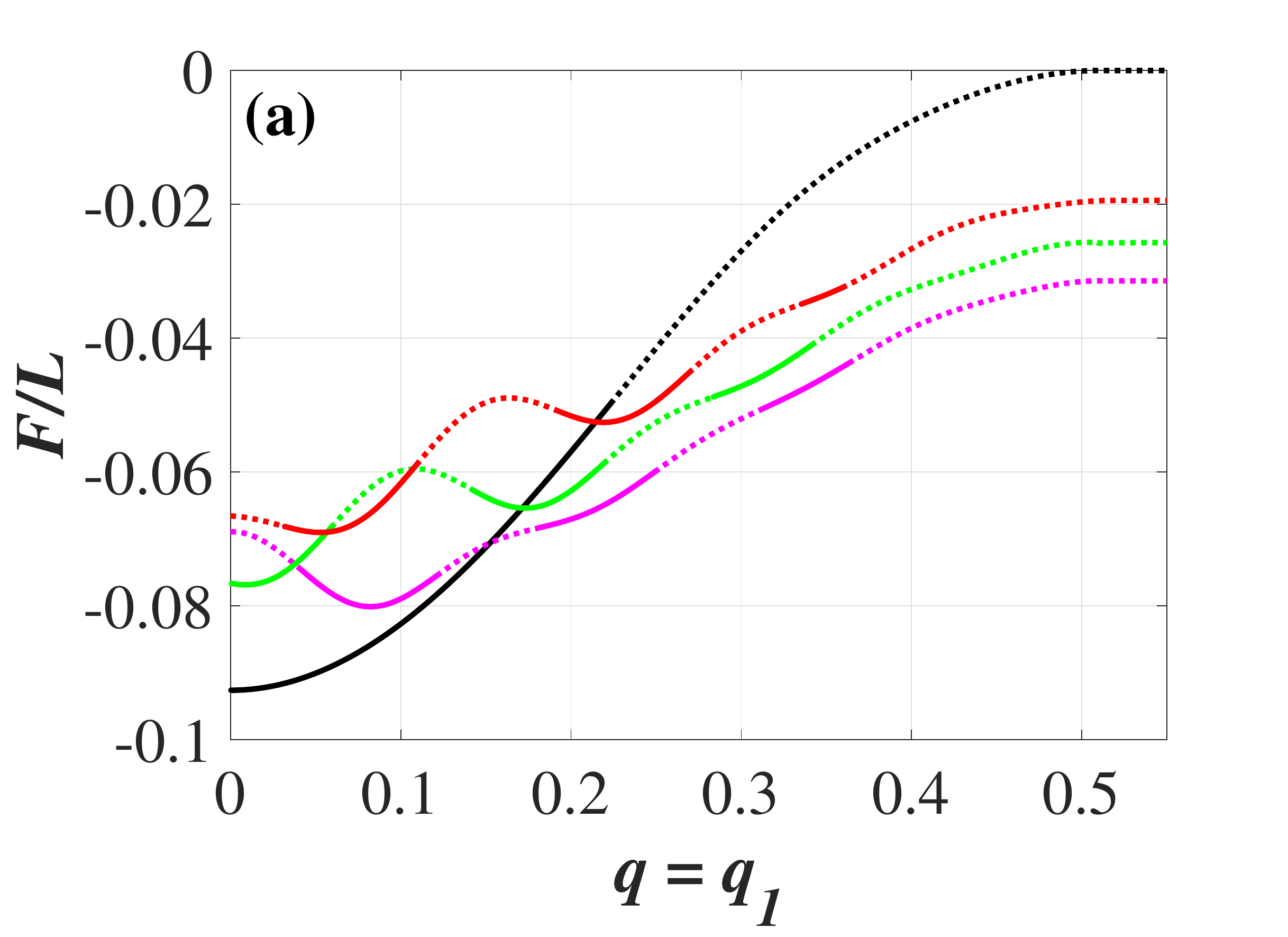}
\includegraphics[width=0.32\columnwidth]{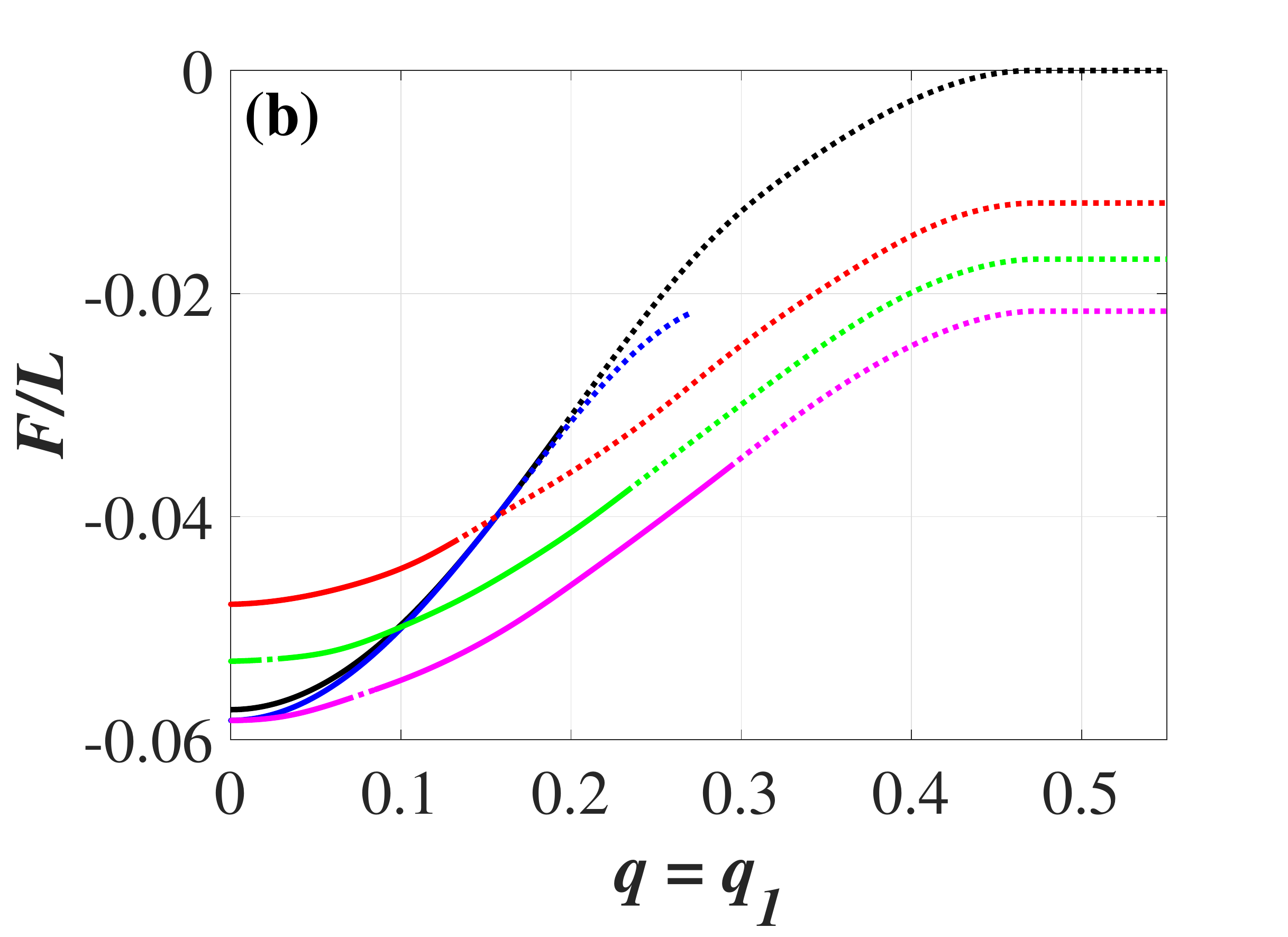}
\includegraphics[width=0.32\columnwidth]{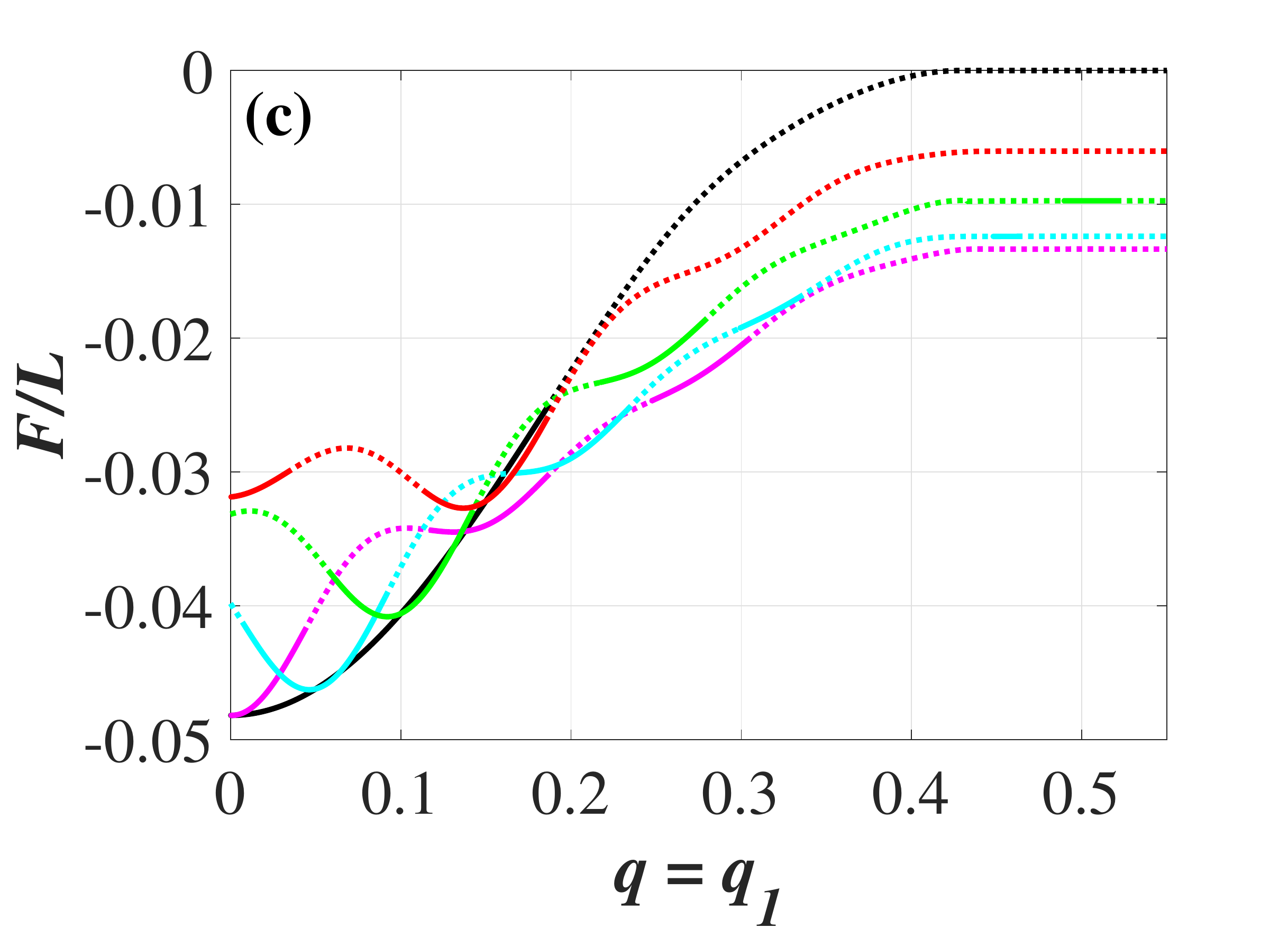}
\caption {GL free energy of a quasi-one-dimensional wire for three reference points as a function of the superfluid momentum $q$ or $q_1$ for the case of the multiple-$q$ state with a given value of $q_2$. In figure (a) the curves are plotted for  $\Gamma=0.07T_{c0}$ and correspond to the non-BTRS state with $s_{\pm}$ symmetry (black line) and multiple-$q$ state with $q_2=0$ (magenta line),  $q_2=0.1$ (green line) and $q_2=0.15$ (red line). In figure (b)  a BTRS state with chiral $s_{\pm}+is_{++}$  symmetry (blue line), a non-BTRS state with $s_{\pm}$ symmetry (black line) and a multiple-$q$ state with $q_2=0$ (magenta line),  $q_2=0.1$ (green line) and $q_2=0.15$ (red line) are depicted for $\Gamma=0.07982T_{c0}$. Figure (c) represents a non-BTRS state with $s_{++}$ symmetry (black line) and a multiple-$q$ state with  $q_2=0$ (magenta line),  $q_2=0.05$ (cyan line), $q_2=0.1$ (green line) and $q_2=0.15$ (red line) when $\Gamma=0.09T_{c0}$. Solid lines for all curves refer to stable regions of the above mentioned states, while dotted lines indicate saddle or unstable regions. The ratio of diffusion coefficients $D_2/D_1 = 2$.}
\label{GL_energy_plots}
\end{figure}

As noted in the introduction, the emergence of additional degrees of freedom of the order parameter give rise to a plenty of current states in multicomponent superconducting systems. In the case of two components, we have already seen that the superconducting momenta of the Cooper pairs of each component admit both equal and different values, forming at least two homogeneous (BTRS and non-BTRS) and one inhomogeneous (multiple-$q$) state. The cornerstone for understanding the mechanisms of possible switching and phase transitions between these states is the behavior of the GL free energy depending on the superconducting momentum $q_1$ or momenta $q_1$, $q_2$. Obviously, this can be done by solving the equations for the order parameters derived for each state and then substituting them into the expressions for the GL energy of the superconducting wire. 

To this end we would like to emphasize that for the BTRS state we use Eqs.\ (\ref{OP1_sol1q}), (\ref{OP2_sol2q}) and (\ref{Gibbs_final_BTRS}); for the non-BTRS state governing equations are Eqs.\ (\ref{GLeq1_homo}), (\ref{GLeq1_homo}) and (\ref{GL_homo}). Finally, the calculations of the multiple-$q$ state exploit the Eqs.\ (\ref{GLeq1_q1-q2}), (\ref{GLeq2_q1-q2}) and (\ref{GL_q1-q2}). For the latter case we need to choose and fix certain values of $q_2$ and consider $q_1$ as the $q$ variable for the BTRS and non-BTRS state.  Such a trick allows to compare the energies of multiple-$q$ state, characterized by two superfluid momenta $q_1$ and $q_2$, with energies of homogeneous states with the unique superfluid momentum $q$. In order to exclude the Josephson effect, we consider a very long channel with a length exceeding the coherence lengths and the London penetration depths for each component of the order paraneter. From the numerical point of view here and hereafter, we set the channel length is equal to  $L=50\xi_{10}$, where $\xi_{10}$ is the coherence length for the first component in the absence of the interband interaction at $T=0$.

The results of our calculations are summarized and visualized in Figure \ref{GL_energy_plots}. First of all, in these three figures, corresponding to the reference points selected earlier (see the black filled square, the circle and the diamond in Figure \ref{Phase_diag}), we will identify the energy curves for the homogeneous states, namely BTRS and non-BTRS. The black curves display the non-BTRS state energies as a function of the superconducting momentum $q$ when the phase difference between the order parameters is $\phi = \pi$ ($s_{\pm}$ pairing symmetry) as in Fig. \ref{GL_energy_plots}(a) and (b) or $\phi = 0$  ($s_{++}$ pairing symmetry) as in Fig. \ref{GL_energy_plots}(c). The GL energy behavior of the BTRS state, in which $s_{\pm}+is_{++}$ chiral symmetry occurs, is depicted by the blue line in Fig. \ref{GL_energy_plots}(b). The variety of energy dependences of the inhomogeneous multiple-$q$ state, when the superconducting momenta of the Cooper pairs of each of the two components may differ, is shown by the remaining color curves, where we measure GL energy as a function of $q_1$ (this refinement is additionally shown on the horizontal axis of the graphs as the equality $q=q_1$). The strategy of values selection of momentum $q_2=0$ (magenta line), $q_2=0.05$ (cyan line in Fig. \ref{GL_energy_plots}(c) only), $q_2=0.1$ (green line) and $q_2=0.1$ (red line) is arbitrary and was due solely to the demonstration of the variety of transitions, which will be discussed below. In other words, without loss of generality we could choose other values of $q_2$ to compare energies of multiple-$q$ state with its homogeneous counterparts. 

An important detail characterizing the energy behavior of the states is the presence of regions on the curves in figure \ref{GL_energy_plots}, marked with a dotted line, corresponding to the unstable superconducting state. In turn, 
the appearance of such regions is determined by the behavior of the minimal eigenvalues of the Hessian matrix formed by the second partial derivatives of the energy for given values of $q$ (homogeneous state) or values of $q_1$ and $q_2$ (inhomogeneous state). It is well-known that for a function of three or more variables a local minimum is attained, when the Hessian is positive definite, namely has all eigenvalues positive. Therefore, if the minimal eigenvalue is positive then we can make a unambiguous statement about the minimum of the GL energy and as a consequence stability of the given state. We studied this problem in detail in Appendix \ref{sec:C} and after that specified the regions of instability as dotted lines. 

From this additional elucidation stems the full picture of possible phase transitions between homogeneous and inhomogeneous state of a dirty two-component superconductor. As a starting point, we consider how a system evolves where at $q=q_1=0$ the ground state is the non-BTRS state with $s_{++}$ pairing symmetry (see Fig. \ref{GL_energy_plots}a). With increasing value $q$ the system moves on the energy scale along the black curve denoting a homogeneous state. Figure  \ref{GL_energy_plots}a shows that at a certain value of $q=q_1$ it crosses the dotted magenta energy line, which, however, is unstable and thus cannot transit to this inhomogeneous state with $q_2=0$. As a consequence, with a further increase in $q$, when the black curve crosses already with the green solid curve, there is a transition to the multiple-$q$ state with $q_2=0.1$. The system energy then evolves along the green curve until it attains the next unstable region (the green dotted curve). After that, one can say that either the system stabilizes here or descends to the lower energy level (magenta curve), where it continues its evolution moving along this curve to the unstable area. 
The scenario described is obviously a probabilistic one, since the arbitrary character of the choice of the $q_2$ values for our  energy plots was already mentioned above. In this particular example we have only demonstrated how this inhomogeneous multiple-$q$ state can emerge in a superconducting wire. 

To describe the evolution of current states and phase transitions between them in Figure  \ref{GL_energy_plots}, we should note first that even without any multiple-$q$ state, there is a direct possibility of of the first order phase transition between BTRS and non-BTRS states with the increase of $q$, when the blue curve (BTRS state) meets the black curve (non-BTRS state). The existence of such a topological transition has already been predicted in the case of systems with the Euler characteristic equal to zero (double-connected systems of the cylinder or ring type etc \cite{Yerin_Cuoco}. Now it can be seen that this prediction can be extended to the case of a quasi-dimensional channel as well. 

The account of the multiple-$q$ state adds essential features to the evolutionary processes of the system under consideration.  The most remarkable feature in this case is the coincidence (within the numerical error of calculations) of the
 energies at $q=0$ for the homogeneous BTRS (blue line) and the inhomogeneous multiple-$q$ (magenta line) states and, as a result, the possibility for the system with equal probabilities to evolve by two different paths with increasing $q$. The first path is the choice and motion of the system within the multiple-$q$ state with $q_2=0$ as long as that state remains stable (solid line). The second one represents the evolution as the homogeneous BTRS state (blue line) with the subsequent transition to the inhomogeneous  multiple-$q$ state with $q_2=0.1$ (green line), which in turn, as $q=q_1$ increases, can exist within the stable region and then may relax into already known inhomogeneous state with $q_2=0$ (magenta line) that is favorable from an energetic point of view.

As for Figure  \ref{GL_energy_plots}c and the probable scenario of the evolution of the current states, the picture looks even richer and more diverse with its phase transitions due to the chosen values of $q_2$. First, as in the previous case in Figure  \ref{GL_energy_plots}b, the energies of the homogeneous non-BTRS state (black line) and inhomogeneous multiple-$q$ (magenta line) states with $q_2=0$ at $q=q_1=0$ coincide (within the accuracy of our numerical calculation). With increasing $q=q_1$ this allows the system to start to evolve equally probable both these states. Second, regardless of the initial state with increasing $q_1$ the system can undergo a cascade of transitions. For instance, let us consider the non-BTRS state (black line) as the starting stage of current states. One can easily see that an increase in momentum $q=q_1$ is accompanied by a switch of the non-BTRS state to the multiple-$q$ state with $q_2=0.05$ (cyan line). Then the system comes back to the homogeneous non-BTRS state (black line). After that the transition to another multiple-$q$ state with $q_2=0.1$ occurs (green line) and a fall again to the non-BTRS state. Finally, this cascade completes by the transition from the non-BTRS state to the multiple-$q$ state with $q_2=0$ (magenta line), where further evolution is restricted by the condition of stability (dotted magenta line corresponds to the unstable state).

\section{Discussions}
\label{sec:Discussion}
\begin{figure}
\includegraphics[width=0.32\columnwidth]{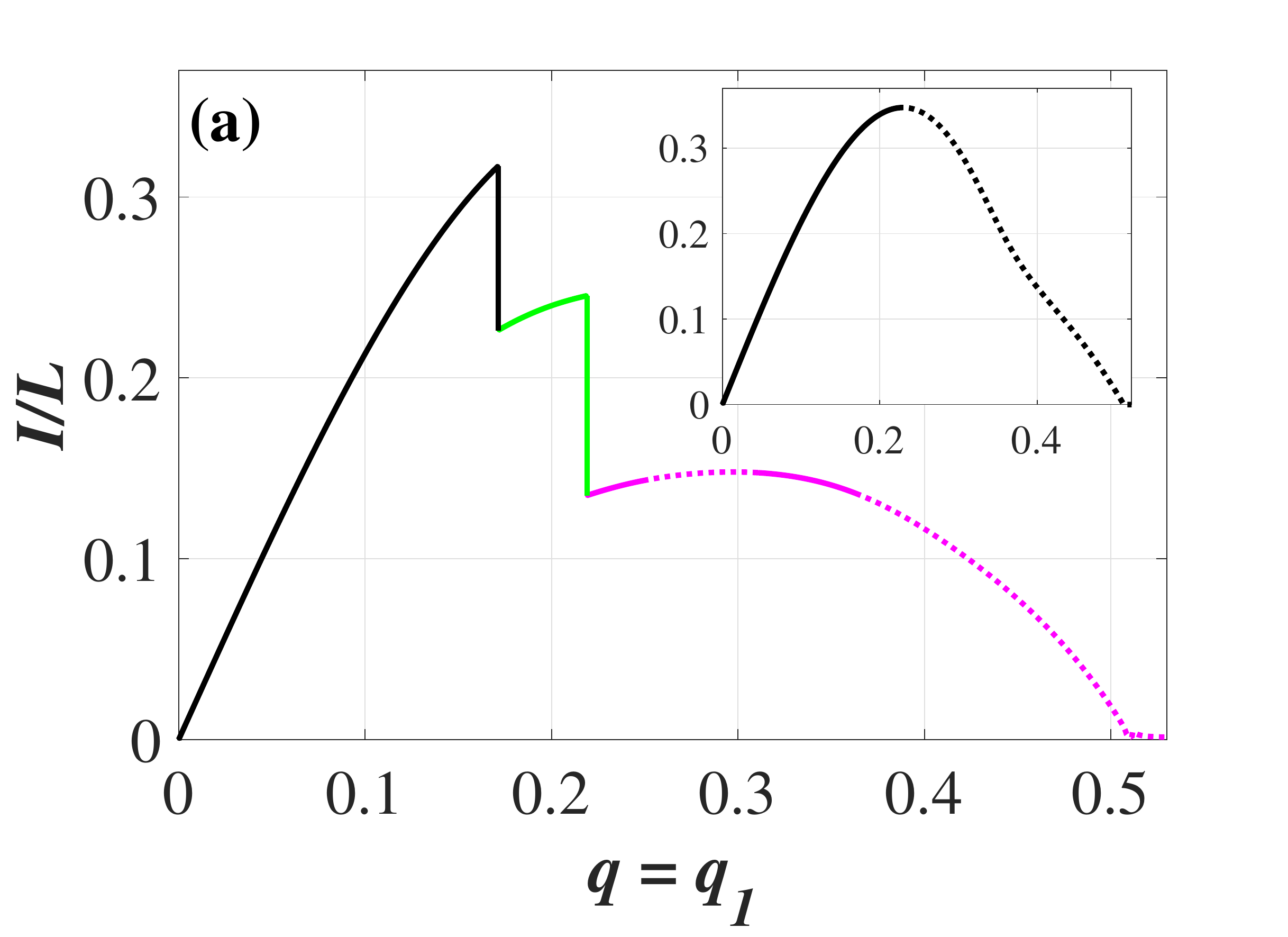}
\includegraphics[width=0.32\columnwidth]{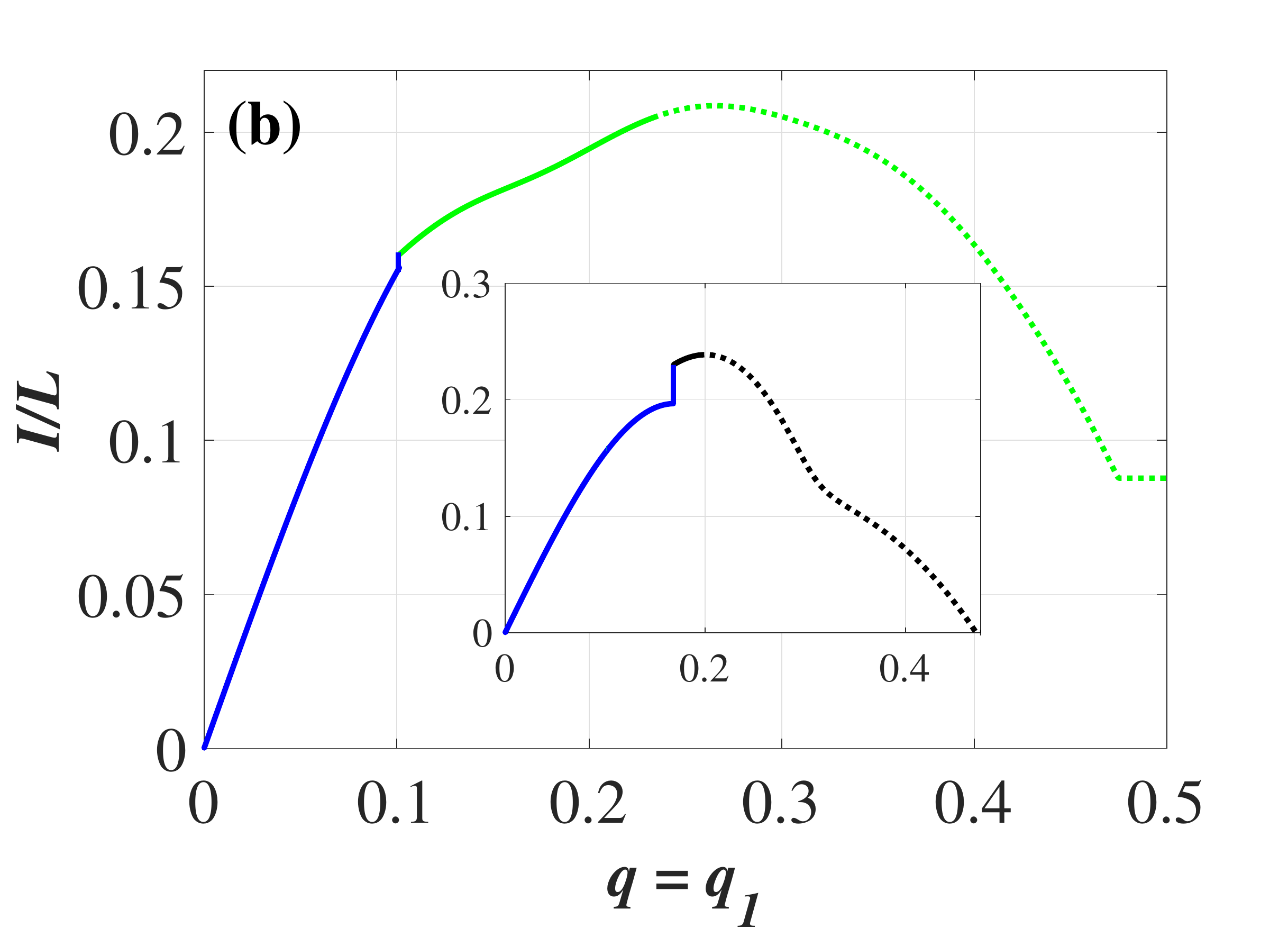}
\includegraphics[width=0.32\columnwidth]{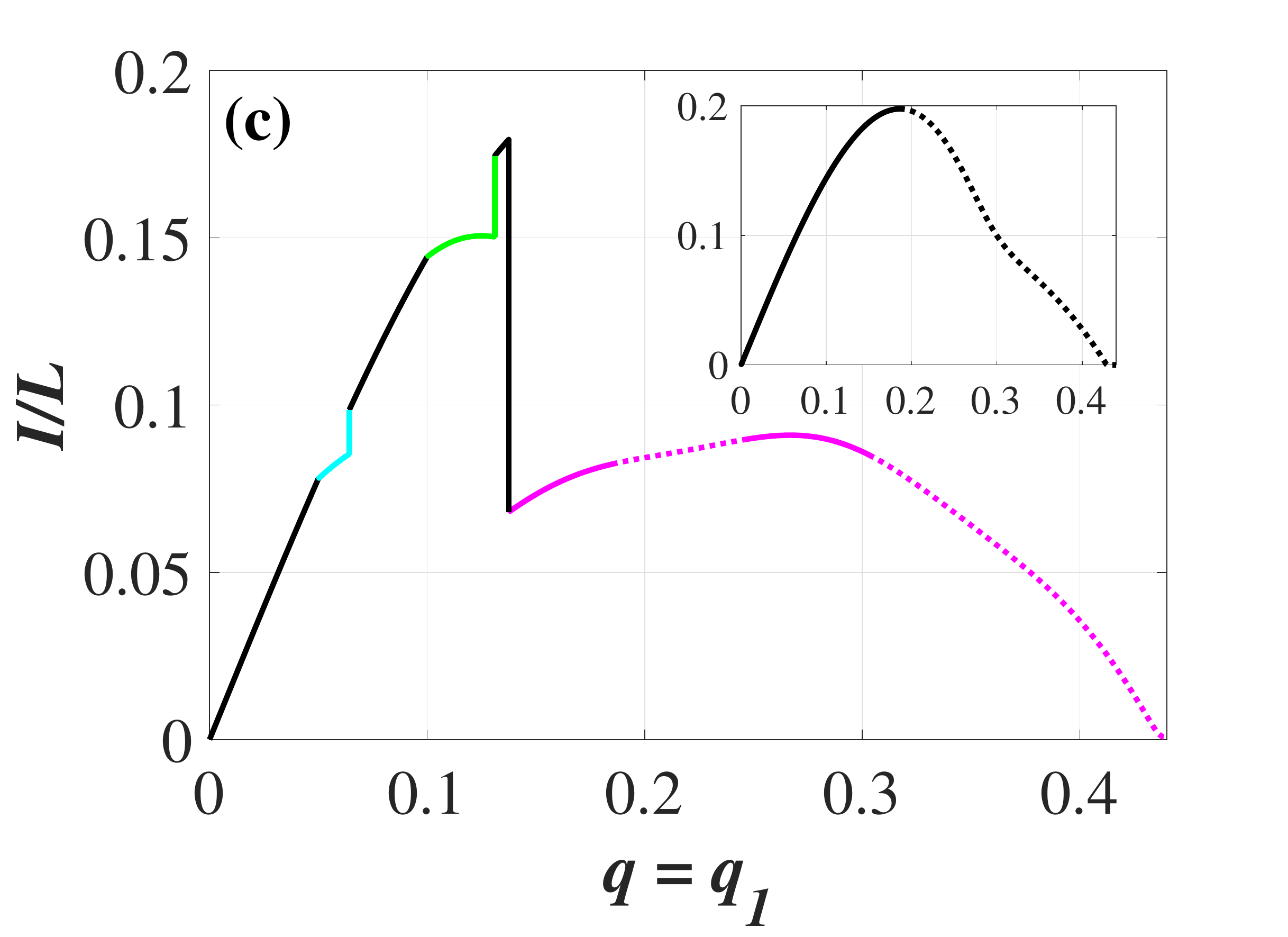}
\caption {Patterns of possible dependencies of the total current $I$ in a quasi-one-dimensional wire for three reference points vs.\ the superfluid momenta $q$ or $q_1$ for the case of a multiple-$q$ state with a given value of $q_2$. (a) For the non-BTRS case with $s_{\pm}$ symmetry ($\Gamma=0.07T_{c0}$) the current-momentum dependence consists of the contribution from the $\phi=\pi$ state (black line) and multiple-$q$ states with $q_2=0.1$ (green line) and $q_2=0$ (magenta line). The inset in (a) shows the current-momentum dependence for a non-BTRS state without phase transitions between different states. (b) For the BTRS case ($\Gamma=0.07982T_{c0}$) the current-momentum dependence can be formed by contributions by the chiral state with $\phi \ne 0$ (blue line) and  from multiple-$q$ state with $q_2=0$ (magenta line). In the absence of a multiple-$q$ state the current-momentum dependence has the form shown in the inset where the black curve is for the non-BTRS state with  $\phi=\pi$ ($s_{\pm}$ symmetry). (c) The phase transitions between the non-BTRS state with $s_{++}$ symmetry ($\phi=0$) and multiple-$q$ state can be detected by the current-momentum dependence, which may be formed by contributions of the $\phi=0$ state (black line) and multiple-$q$ states with $q_2=0.05$ (cyan line), $q_2=0.1$ (green line) and the $q_2=0$ (magenta line). The inset in (c) shows the current-momentum dependence for a "pure" non-BTRS state without transitions between different states. In all figures the solid and dotted lines of corresponding colors specify the stable and unstable states, respectively. The ratio of diffusion coefficients $D_2/D_1 = 2$.}
\label{current_plots}
\end{figure}

The existence of such transitions obviously raises the question of how to record them experimentally or outline possible experimental strategies for detecting them. One possibly suitable method of their observation is the study of the so-called depairing current (current-momentum) curves, related to transport properties. In other words one should  measure the dependences of the depairing current at which the kinetic energy of the superconducting carriers equals the binding energy of the Cooper pairs, i.e.\ when the value of the current reaches the certain threshold above which superconductivity is suppressed. Such dependences can be calculated based on Eqs. (\ref{current_BTRS}),  (\ref{current_homo}) and (\ref{current_q1q2}) taking into account dependences of energy (see Fig. (\ref{GL_energy_plots}) with possible scenarios for the evolution of states as shown in the previous section. 

Using such a probe one can plot the depairing currents to show the hallmarks of phase transitions between different states, in particular between homogeneous BTRS or non-BTRS state and inhomogeneous multiple-$q$ state (Fig. \ref{current_plots}). 

The interesting feature worth mentioning is the presence of two stable increasing with $q$  branches of the depairing curve corresponding to the possibility of a bistable state (see the solid lines in Fig. \ref{current_plots}a and c). The dotted regions of the depairing curves, as in the case of the energy dependences, display unstable regions that will be not observed during an experiment. They do not carry any physical meaning and thus cannot be measured. The same conclusion applies to the plateau of the depairing curve at large values of $q=q_1$ and the non-zero value of $q_2=0.1$ for the BTRS state case, shown by the dotted line in Figure \ref{current_plots}b. This result is an artifact of our assumption of a initially fixed $q_2 \ne 0$ and does not reflect the real transport properties of the system belonging to instability of the superconducting phase.

Moreover, for conventional superconductors, it has long ago been established that the monotonically increasing part of the depairing curve corresponds to a stable superconducting state, while the monotonically decreasing part corresponds to an instability.  A remarkable counterintuitive feature found here is that within the multiple-$q$ state the depairing current curves exhibit an increasing segment, which can be unexpectedly essentially unstable (see dotted magenta lines in Fig. \ref{current_plots}a and c and the dotted green line in Fig. \ref{current_plots}b), too. And vice versa, there are decreasing segments corresponding stable states (see solid magenta lines in Fig. \ref{current_plots}a and c).

It should be noted that the plots are illustrative in nature and are intended to demonstrate the expected noteworthy qualitative characteristics of phase transitions between homogeneous and inhomogeneous states. 

From a measurement perspective the experimental verification of predicted results can be done by means of so-called pulsed measurement technique, which has already proven itself in the study of superconducting transport properties and the detection of the depairing current in particular. A technical description of this experimental approach and further details on the depairing current can be found elsewhere (see e.g.\ \onlinecite{Kunchur1, Kunchur2, Liang2013, Matsushita2019}).

\section{Conclusions}
\label{sec:Conclusion}
In this paper, using the Ginzburg-Landau theory for a two-band superconductor with the interband impurity scattering effect as the underlying model, we have extended the variety of exotic states in multicomponent superconductors. For a quasi-one-dimensional thin wire or channel we predict the emergence of an inhomogeneous multiple-$q$ state, which is characterized by different superconducting condensate momenta. Based on a particular example of a two-band superconductor with the weak repulsive interband interaction, we have revealed that the multiple-$q$ state can trigger a peculiar cascade of phase transitions between this novel state and the homogeneous BTRS or non-BTRS states and vice versa. A possibly suitable tool for the detection of this inhomogeneous state has been proposed to verify our theoretical predictions.  According to our calculations, a saw-like dependence of the depairing current and the emergence of the bistable current state can be considered as a fingerprint of such a multiple-$q$ state. A quantitative comparison with the two $q$-vectors expected in the FFLO state in the bulk is an interesting issue for future studies. Both phenomena are expected to shed light on the rich response of unconventional multiband superconductors and the complexity of their pair condensates.

\begin{acknowledgments}
Y.Y.\ acknowledges support by the CarESS project. 
\end{acknowledgments}

\begin{widetext}
\appendix
\section{GL coefficients}
\label{sec:A}

The coefficients of the GL free energy functional Eq. (\ref{GL_general}) are expressed as \cite{Stanev, Corticelli}:
\begin{equation}
\label{a_i}
{a_{ii}} = {N_i}\left( {\frac{{{\lambda _{jj}}}}{{\det {\lambda _{ij}}}} - 2\pi T\sum\limits_{\omega  > 0}^{{\omega _c}} {\frac{{\omega  + {\Gamma _{ij}}}}{{\omega \left( {\omega  + {\Gamma _{ij}} + {\Gamma _{ji}}} \right)}}} } \right) = {N_i}\left( {\frac{{{\lambda _{jj}}}}{{\det {\lambda _{ij}}}} - \frac{1}{\lambda } + \ln \left( {\frac{T}{{{T_c}}}} \right) + \psi \left( {\frac{1}{2} + \frac{\Gamma }{{\pi T}}} \right) - \psi \left( {\frac{1}{2}} \right)} \right),
\end{equation}
\begin{equation}
\label{a_12}
{a_{ij}} =  - {N_i}\left( {\frac{{{\lambda _{ij}}}}{{\det {\lambda _{ij}}}} + 2\pi T\sum\limits_{\omega  > 0}^{{\omega _c}} {\frac{{{\Gamma _{ij}}}}{{\omega \left( {\omega  + {\Gamma _{ij}} + {\Gamma _{ji}}} \right)}}} }, \right)
\end{equation}
\begin{equation}
\label{b_i}
{b_{ii}} = {N_i}\pi T\sum\limits_{\omega  > 0}^{{\omega _c}} {\frac{{{{\left( {\omega  + {\Gamma _{ji}}} \right)}^4}}}{{{\omega ^3}{{\left( {\omega  + {\Gamma _{ij}} + {\Gamma _{ji}}} \right)}^4}}}}  + {N_i}\pi T\sum\limits_{\omega  > 0}^{{\omega _c}} {\frac{{{\Gamma _{ij}}\left( {\omega  + {\Gamma _{ji}}} \right)\left( {{\omega ^2} + 3\omega {\Gamma _{ji}} + \Gamma _{ji}^2} \right)}}{{{\omega ^3}{{\left( {\omega  + {\Gamma _{ij}} + {\Gamma _{ji}}} \right)}^4}}}},
\end{equation}
\begin{equation}
\label{b_12}
{b_{ij}} =  - {N_i}\pi T\sum\limits_{\omega  > 0}^{{\omega _c}} {\frac{{{\Gamma _{ij}}{\omega ^3}}}{{{\omega ^3}{{\left( {\omega  + {\Gamma _{ij}} + {\Gamma _{ji}}} \right)}^4}}}}  + {N_i}\pi T\sum\limits_{\omega  > 0}^{{\omega _c}} {\frac{{{\Gamma _{ij}}\left( {{\Gamma _{ij}} + {\Gamma _{ji}}} \right)\left( {{\Gamma _{ji}}\left( {\omega  + 2{\Gamma _{ij}}} \right) + \omega {\Gamma _{ij}}} \right)}}{{{\omega ^3}{{\left( {\omega  + {\Gamma _{ij}} + {\Gamma _{ji}}} \right)}^4}}}},
\end{equation}
\begin{equation}
\label{c_i}
{c_{ii}} = {N_i}\pi T\sum\limits_{\omega  > 0}^{{\omega _c}} {\frac{{{\Gamma _{ij}}\left( {\omega  + {\Gamma _{ji}}} \right)\left( {{\omega ^2} + \left( {\omega  + {\Gamma _{ji}}} \right)\left( {{\Gamma _{ij}} + {\Gamma _{ji}}} \right)} \right)}}{{{\omega ^3}{{\left( {\omega  + {\Gamma _{ij}} + {\Gamma _{ji}}} \right)}^4}}}},
\end{equation}
\begin{equation}
\label{c_12}
{c_{ij}} = {N_i}\pi T\sum\limits_{\omega  > 0}^{{\omega _c}} {\frac{{{\Gamma _{ij}}\left( {\omega  + {\Gamma _{ji}}} \right)\left( {\omega  + {\Gamma _{ji}}} \right)\left( {{\Gamma _{ij}} + {\Gamma _{ji}}} \right)}}{{{\omega ^3}{{\left( {\omega  + {\Gamma _{ij}} + {\Gamma _{ji}}} \right)}^4}}}},
\end{equation}
\begin{equation}
\label{k_i}
{k_{ii}} = 2{N_i}\pi T\sum\limits_{\omega  > 0}^{{\omega _c}} {\frac{{{D_i}{{\left( {\omega  + {\Gamma _{ji}}} \right)}^2} + {\Gamma _{ij}}{\Gamma _{ji}}{D_j}}}{{{\omega ^2}{{\left( {\omega  + {\Gamma _{ij}} + {\Gamma _{ji}}} \right)}^2}}}}
\end{equation}
\begin{equation}
\label{k_12}
{k_{ij}} = 2{N_i}{\Gamma _{ij}}\pi T\sum\limits_{\omega  > 0}^{{\omega _c}} {\frac{{{D_i}\left( {\omega  + {\Gamma _{ji}}} \right) + {D_j}\left( {\omega  + {\Gamma _{ij}}} \right)}}{{{\omega ^2}{{\left( {\omega  + {\Gamma _{ij}} + {\Gamma _{ji}}} \right)}^2}}}},
\end{equation}
where $\omega=(2n+1)\pi T$  are Matsubara frequencies, $\omega_c$ is the cut-off frequency,  $N_i$ are the densities of states at the Fermi level, $\lambda_{ij}$  and $\Gamma_{ij}$ are coupling constants and interband scattering rates that characterize the strength of the interband impurities, $D_i$ are diffusion coefficients. For the sake of simplicity and without loss of generality we put $\lambda_{12}=\lambda_{21}$, $\Gamma_{12}=\Gamma_{21}$ and $N_1=N_2$ in the main paper.

Eqs.\ (\ref{a_12})-(\ref{k_12}) can be expressed in terms of polygamma functions after the summation procedure. However, we do not provide these expression due to their cumbersome forms. 

\section{The critical temperature as a function of impurities and the strength of the interband interaction}
\label{sec:B}

The expression for the critical temperature as a function of the impurity scattering rate $\Gamma$ can be obtained within the linearized Usadel equations generilized for two-band superconductor and supplemented by the self-consistent equations for the energy gaps (see details in Ref.\ \onlinecite{Gurevich1}). The final formula showing the suppression of the critical temperature $T_c$ in respect to the critical temperature $T_{c0}$ of a clean two-band superconductor without impurities when $\Gamma=0$ is given by
\begin{equation}
\label{T_c_Usadel}
U\left( {\frac{\Gamma }{{\pi T_c}}} \right) =  - \frac{{2\left( {w\lambda \ln t + \lambda \left( {{\lambda _{11}} + {\lambda _{22}}} \right) - 2w} \right)\ln t}}{{2w\lambda \ln t + \lambda \left( {{\lambda _{11}} + {\lambda _{22}} - {\lambda _{12}} - {\lambda _{21}}} \right) - 2w}},
\end{equation}
where $U\left( x \right) = \psi \left( {\frac{1}{2} + x} \right) - \psi \left( {\frac{1}{2}} \right)$ is expressed via  the digamma function $\psi(x)$, $t=T_c/T_{c0}$, $\lambda$ is the largest eigenvalue of the matrix of intra- and interband coefficients and $w = \det {\lambda _{ij}} = {\lambda _{11}}{\lambda _{22}} - {\lambda _{12}}{\lambda _{21}}$.
\begin{figure}
\includegraphics[width=0.49\columnwidth]{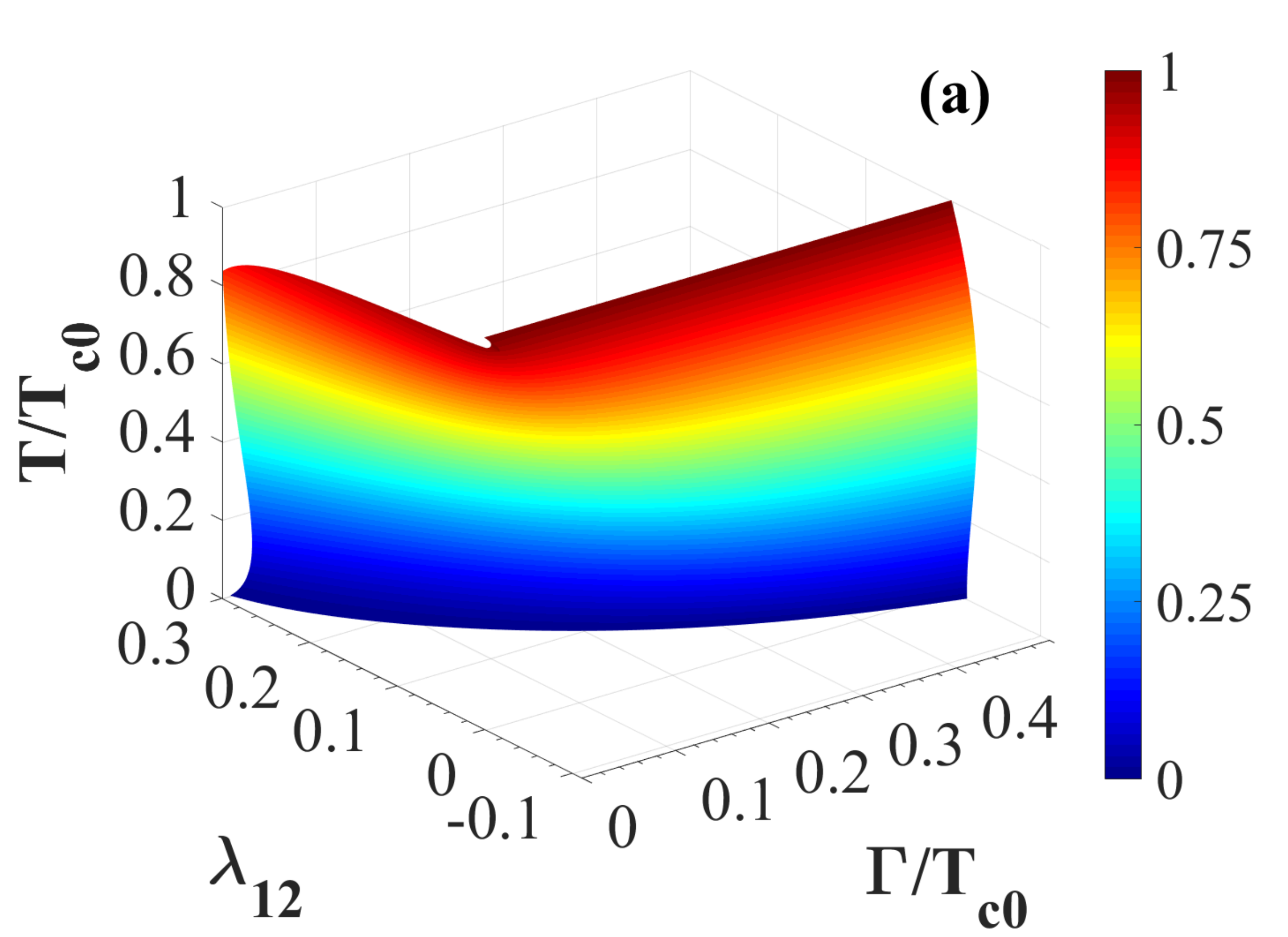}
\includegraphics[width=0.49\columnwidth]{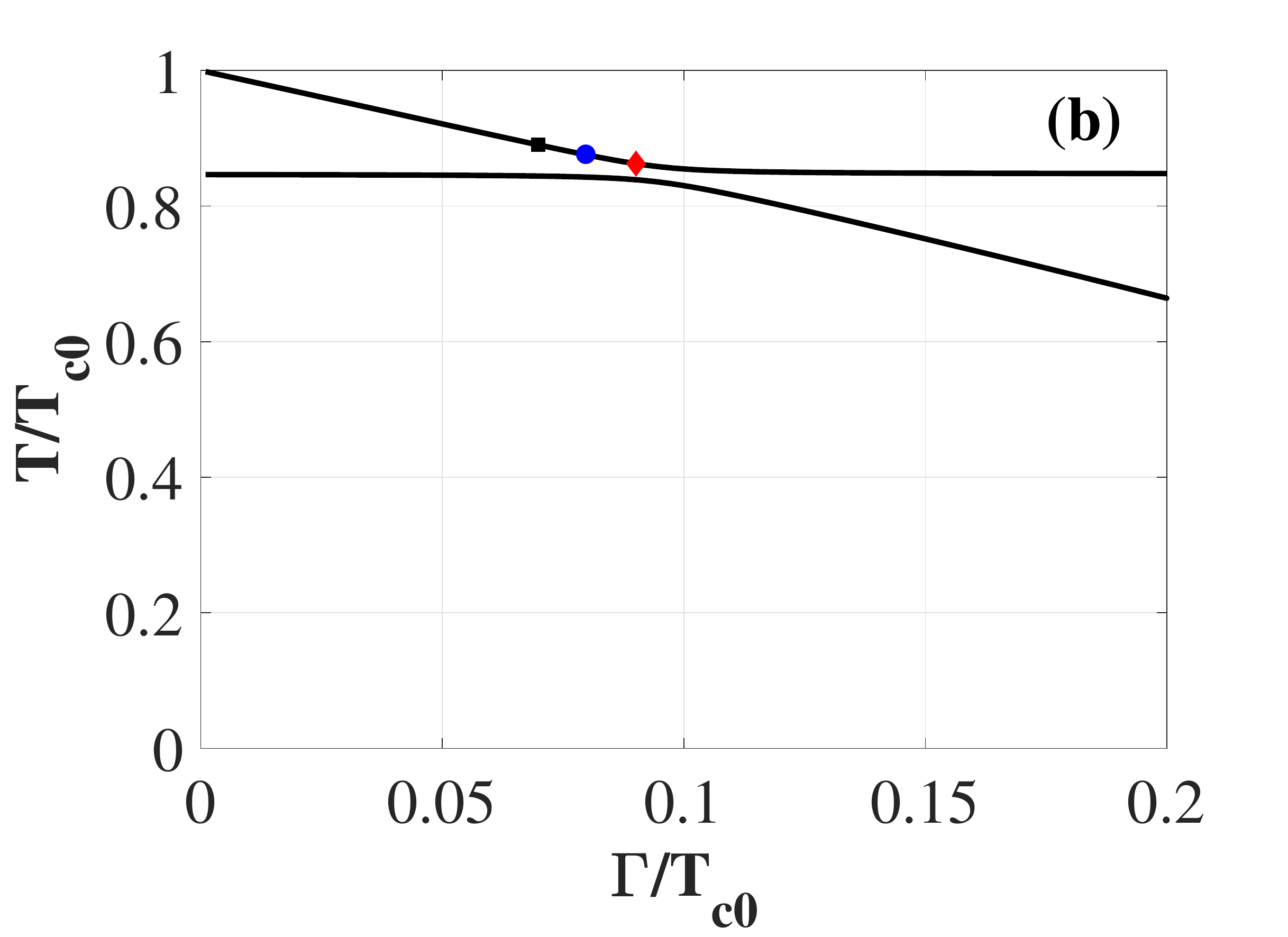}
\caption {(a) The critical temperature $T_c$ of a dirty two-band superconductor as a function of the interband scattering rate $\Gamma$ and the interband interasction coefficient $\lambda_{12}$ with $\lambda_{11}=0.35$ and $\lambda_{22}=0.347$ . (b) $T_c$ as a function of $\Gamma$ with $\lambda_{11}=0.35$, $\lambda_{22}=0.347$, $\lambda_{12}=\lambda_{21}=-0.01$. The values of $T_c$ and $\Gamma$ are calibrated to the critical temperature of a two-band superconductor without impurities $T_{c0}$ and $\Gamma = 0$, respectively. The filled black square, blue circle and red diamond correspond to values of $\Gamma=0.07T_{c0}$, $\Gamma=0.07982T_{c0}$ and $\Gamma=0.09T_{c0}$, which are considered in the main paper as reference points for $s_{\pm}$ (non-BTRS state), $s_{\pm}+is_{++}$ (BTRS state) and $s_{++}$ (non-BTRS state) simmetries of the order parameter. The low black cirve in (b) represents unphysical solution of Eq. (\ref{T_c_Usadel}).}
\label{Tc_vs_Gamma}
\end{figure}

The numerical solution of Eq.\ (\ref{T_c_Usadel}) is shown in Figure \ref{Tc_vs_Gamma}. It is interesting that generally speaking Eq. (\ref{T_c_Usadel}) has two type of solutions, one of them (lower curve in Figure \ref{Tc_vs_Gamma}b) is unphysical. To make it more convincing we have marked the filled black square, blue circle and red diamond on the upper curve as the reference points presented earlier on the phase diagram in Figure \ref{Phase_diag}.

\section{Stability conditions}
\label{sec:C}
\subsection{The BTRS state}
\begin{figure}
\includegraphics[width=0.5\columnwidth]{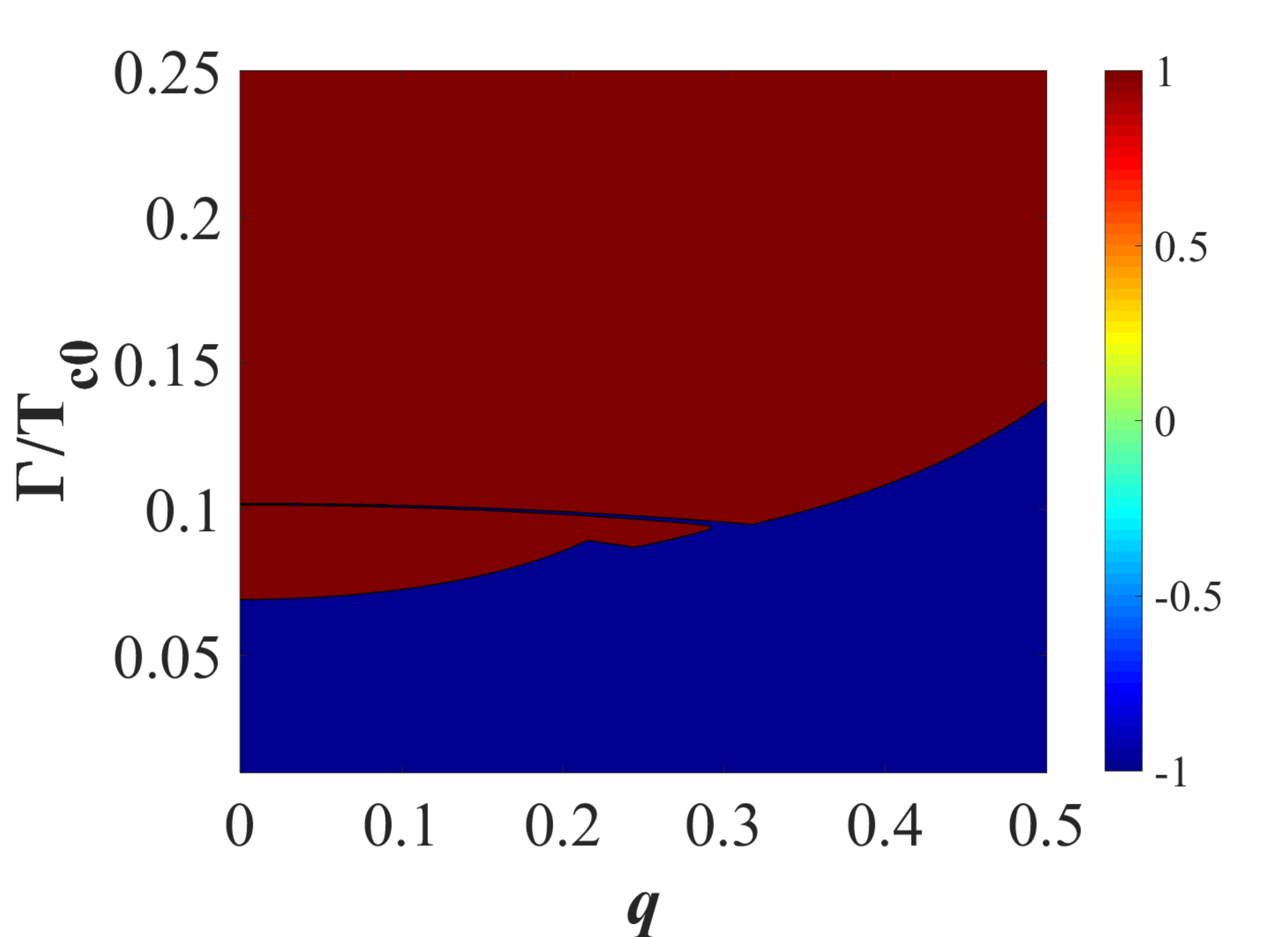}
\caption {The contour plot of the minimal eigenvalue of the Hessian matrix as a function of $q$ and $\Gamma$ for the BTRS state. The minimum of the GL energy has been found for the red region. The blue region might correspond to saddle-points or maxima.}
\label{stability_BTRS}
\end{figure}
The problem of the current state stability is equivalent to the problem of determining the point of extrema of the GL free energy as a minimum, maximum or saddle point. Since in the case of the BTRS state Eq. (\ref{Gibbs_final_BTRS}) is a function of three variables $\Delta_1$, $\Delta_2$ and $q$, the problem is reduced to the study of the eigenvalues of the Hessian matrix at the critical point. The Hessian matrix has the form
\begin{equation}
\label{Hessian_BTRS}
{H_{\left| {{\Delta _1}} \right|\left| {{\Delta _1}} \right|q}} = \left( {\begin{array}{*{20}{c}}
  {\frac{{{\partial ^2}F}}{{\partial {{\left| {{\Delta _1}} \right|}^2}}}}&{\frac{{{\partial ^2}F}}{{\partial \left| {{\Delta _1}} \right|\left| {{\Delta _2}} \right|}}}&{\frac{{{\partial ^2}F}}{{\partial \left| {{\Delta _1}} \right|q}}} \\ 
  {\frac{{{\partial ^2}F}}{{\partial \left| {{\Delta _2}} \right|\left| {{\Delta _1}} \right|}}}&{\frac{{{\partial ^2}F}}{{\partial {{\left| {{\Delta _2}} \right|}^2}}}}&{\frac{{{\partial ^2}F}}{{\partial \left| {{\Delta _2}} \right|q}}} \\ 
  {\frac{{{\partial ^2}F}}{{\partial q\left| {{\Delta _1}} \right|}}}&{\frac{{{\partial ^2}F}}{{\partial q\left| {{\Delta _2}} \right|}}}&{\frac{{{\partial ^2}F}}{{\partial {q^2}}}} 
\end{array}} \right),
\end{equation}
where 
\begin{equation}
\frac{{{\partial ^2}F}}{{\partial {{\left| {{\Delta _1}} \right|}^2}}} = \frac{{6{\mkern 1mu} {{\left| {{\Delta _1}} \right|}^2}\left( {{b_{11}}{\mkern 1mu} {c_{12}} - c_{11}^2} \right) + 2{{\left| {{\Delta _2}} \right|}^2}\left( {{b_{12}}{\mkern 1mu} {c_{12}} - {c_{11{\mkern 1mu} }}{c_{22}} - c_{12}^2} \right){\mkern 1mu}  + \left( {{k_{11}}{c_{12}}{\mkern 1mu}  - {k_{12}}{c_{11}}} \right){\mkern 1mu} {q^2} + 2{\mkern 1mu} \left( {{a_{11}}{\mkern 1mu} {c_{12}} - {a_{12}}{\mkern 1mu} {c_{11}}} \right)}}{{{c_{12}}}},
\end{equation}
\begin{equation}
\frac{{{\partial ^2}F}}{{\partial \left| {{\Delta _1}} \right|\left| {{\Delta _2}} \right|}} = \frac{{4{\mkern 1mu} \left| {{\Delta _1}} \right|\left| {{\Delta _2}} \right|{\mkern 1mu} \left( {{b_{12}}{\mkern 1mu} {c_{12}} - {c_{11{\mkern 1mu} }}{c_{22}} - c_{12}^2} \right)}}{{{c_{12}}}},
\end{equation}
\begin{equation}
\frac{{{\partial ^2}F}}{{\partial \left| {{\Delta _1}} \right|q}} = \frac{{{\mkern 1mu} 2\left| {{\Delta _1}} \right|\left( {{c_{12}}{\mkern 1mu} {k_{11}} - {k_{12}}{c_{11}}} \right)q}}{{{c_{12}}}},
\end{equation}
\begin{equation}
\frac{{{\partial ^2}F}}{{\partial {{\left| {{\Delta _2}} \right|}^2}}} = \frac{{2{\mkern 1mu} {{\left| {{\Delta _1}} \right|}^2}\left( {{b_{12}}{\mkern 1mu} {c_{12}} - {c_{11{\mkern 1mu} }}{c_{22}} - c_{12}^2} \right) + 6{\mkern 1mu} {{\left| {{\Delta _2}} \right|}^2}\left( {{b_{22}}{\mkern 1mu} {c_{12}} - c_{22}^2} \right) + \left( {{k_{22}}{c_{12}} - {k_{12}}{c_{22}}{\mkern 1mu} {\mkern 1mu} } \right){\mkern 1mu} {q^2} + 2\left( {{a_{22}}{\mkern 1mu} {c_{12}} - {\mkern 1mu} {a_{12}}{\mkern 1mu} {c_{22}}} \right){\mkern 1mu} }}{{{c_{12}}}},
\end{equation}
\begin{equation}
\frac{{{\partial ^2}F}}{{\partial \left| {{\Delta _2}} \right|q}} = \frac{{2{\mkern 1mu} \left| {{\Delta _2}} \right|{\mkern 1mu} \left( {{k_{22}}{c_{12}}{\mkern 1mu}  - {k_{12}}{c_{22{\mkern 1mu} }}} \right){\mkern 1mu} q}}{{{c_{12}}}},
\end{equation}
\begin{equation}
\frac{{{\partial ^2}F}}{{\partial {q^2}}} = {\mkern 1mu} \frac{{2{\mkern 1mu} {\mkern 1mu} {{\left| {{\Delta _1}} \right|}^2}\left( {{k_{11}}{c_{12}}{\mkern 1mu}  - {k_{12}}{c_{11}}{\mkern 1mu} } \right) + 2{\mkern 1mu} {{\left| {{\Delta _2}} \right|}^2}\left( {{k_{22}}{c_{12}} - {k_{12}}{c_{22}}} \right) - 3{\mkern 1mu} k_{12}^2{q^2} - 2{\mkern 1mu} {a_{12}}{\mkern 1mu} {k_{12}}}}{{2{c_{12}}}}.
\end{equation}

To classify the stability region of the BTRS state it is enough to determine the sign of the minimal eigenvalue $l_{min}$ of the Hessian matrix Eq.\ (\ref{Hessian_BTRS}). The contour plot in Figure \ref{stability_BTRS} shows $\operatorname{sgn}(l_{min})$ for different values of the interband scattering rate $\Gamma$ and $q$.

\subsection{The non-BTRS state}
As in the case of the BTRS state for the homogeneous non-BTRS state the Hessian matrix is formed by the second partial derivatives of the GL free energy Eq. (\ref{GL_homo})
\begin{equation}
\label{Hessian_homo}
{H_{\left| {{\Delta _1}} \right|\left| {{\Delta _1}} \right|q}} = \left( {\begin{array}{*{20}{c}}
  {\frac{{{\partial ^2}F}}{{\partial {{\left| {{\Delta _1}} \right|}^2}}}}&{\frac{{{\partial ^2}F}}{{\partial \left| {{\Delta _1}} \right|\left| {{\Delta _2}} \right|}}}&{\frac{{{\partial ^2}F}}{{\partial \left| {{\Delta _1}} \right|q}}} \\ 
  {\frac{{{\partial ^2}F}}{{\partial \left| {{\Delta _2}} \right|\left| {{\Delta _1}} \right|}}}&{\frac{{{\partial ^2}F}}{{\partial {{\left| {{\Delta _2}} \right|}^2}}}}&{\frac{{{\partial ^2}F}}{{\partial \left| {{\Delta _2}} \right|q}}} \\ 
  {\frac{{{\partial ^2}F}}{{\partial q\left| {{\Delta _1}} \right|}}}&{\frac{{{\partial ^2}F}}{{\partial q\left| {{\Delta _2}} \right|}}}&{\frac{{{\partial ^2}F}}{{\partial {q^2}}}} 
\end{array}} \right),
\end{equation}
with the following expressions for the second derivatives
\begin{equation}
\frac{{{\partial ^2}F}}{{\partial {{\left| {{\Delta _1}} \right|}^2}}} = 2{\mkern 1mu} {a_{11}} + 6{\mkern 1mu} {b_{11}}{\mkern 1mu} {\left| {{\Delta _1}} \right|^2} + 2{\mkern 1mu} {b_{12}}{\mkern 1mu} {\left| {{\Delta _2}} \right|^2} + {k_{11}}{\mkern 1mu} {q^2} + 12{c_{11}}{\mkern 1mu} \left| {{\Delta _1}} \right|{\mkern 1mu} \left| {{\Delta _2}} \right|{\mkern 1mu} {\mkern 1mu} \cos \phi  + 2{\mkern 1mu} {c_{12}}{\mkern 1mu} {\left| {{\Delta _2}} \right|^2}\cos 2\phi,
\end{equation}
\begin{equation}
\frac{{{\partial ^2}F}}{{\partial \left| {{\Delta _1}} \right|\left| {{\Delta _2}} \right|}} = 4{\mkern 1mu} {b_{12}}{\mkern 1mu} \left| {{\Delta _1}} \right|\left| {{\Delta _2}} \right| + {k_{12}}{\mkern 1mu} {q^2} + 2{\mkern 1mu} \left( {{a_{12}} + 3{c_{11}}{\mkern 1mu} {{\left| {{\Delta _1}} \right|}^2} + 3{c_{22}}{{\left| {{\Delta _2}} \right|}^2}} \right)\cos \phi  + 4{\mkern 1mu} {c_{12}}{\mkern 1mu} \left| {{\Delta _1}} \right|{\mkern 1mu} \left| {{\Delta _2}} \right|{\mkern 1mu} \cos 2\phi,
\end{equation}
\begin{equation}
\frac{{{\partial ^2}F}}{{\partial \left| {{\Delta _1}} \right|q}} = 2{\mkern 1mu} q\left( {{k_{11}}\left| {{\Delta _1}} \right|{\mkern 1mu}  + {k_{12}}\left| {{\Delta _2}} \right|{\mkern 1mu} } \right),
\end{equation}
\begin{equation}
\frac{{{\partial ^2}F}}{{\partial {{\left| {{\Delta _2}} \right|}^2}}} = 2{\mkern 1mu} {a_{22}} + 6{\mkern 1mu} {b_{22}}{\mkern 1mu} {\left| {{\Delta _2}} \right|^2} + 2{\mkern 1mu} {b_{12}}{\mkern 1mu} {\left| {{\Delta _1}} \right|^2} + {k_{22}}{\mkern 1mu} {q^2} + 12{c_{22}}{\mkern 1mu} \left| {{\Delta _1}} \right|{\mkern 1mu} \left| {{\Delta _2}} \right|{\mkern 1mu} {\mkern 1mu} \cos \phi  + 2{\mkern 1mu} {c_{12}}{\mkern 1mu} {\left| {{\Delta _1}} \right|^2}\cos 2\phi,
\end{equation}
\begin{equation}
\frac{{{\partial ^2}F}}{{\partial \left| {{\Delta _2}} \right|q}} = 2{\mkern 1mu} q\left( {{k_{12}}\left| {{\Delta _1}} \right|{\mkern 1mu}  + {k_{22}}\left| {{\Delta _2}} \right|{\mkern 1mu} } \right),
\end{equation}
\begin{equation}
\frac{{{\partial ^2}F}}{{\partial {q^2}}} = {\mkern 1mu} {k_{11}}\left| {{\Delta _1}} \right|{{\mkern 1mu} ^2} + 2{k_{12}}{\mkern 1mu} \left| {{\Delta _1}} \right|{\mkern 1mu} \left| {{\Delta _2}} \right|{\mkern 1mu}  + {k_{22}}{\left| {{\Delta _2}} \right|^2}.
\end{equation}

\begin{figure}
\includegraphics[width=0.45\columnwidth]{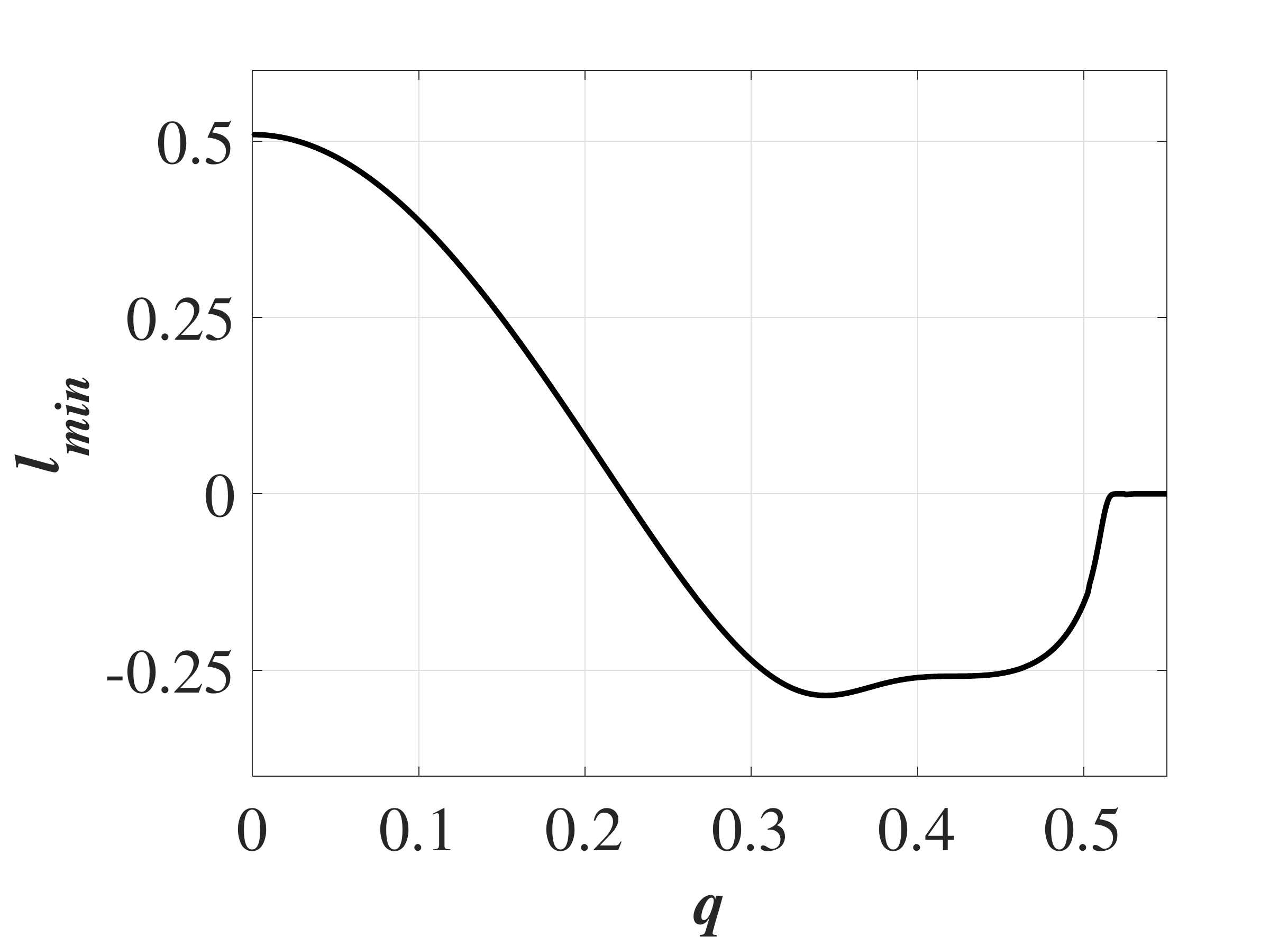}
\includegraphics[width=0.45\columnwidth]{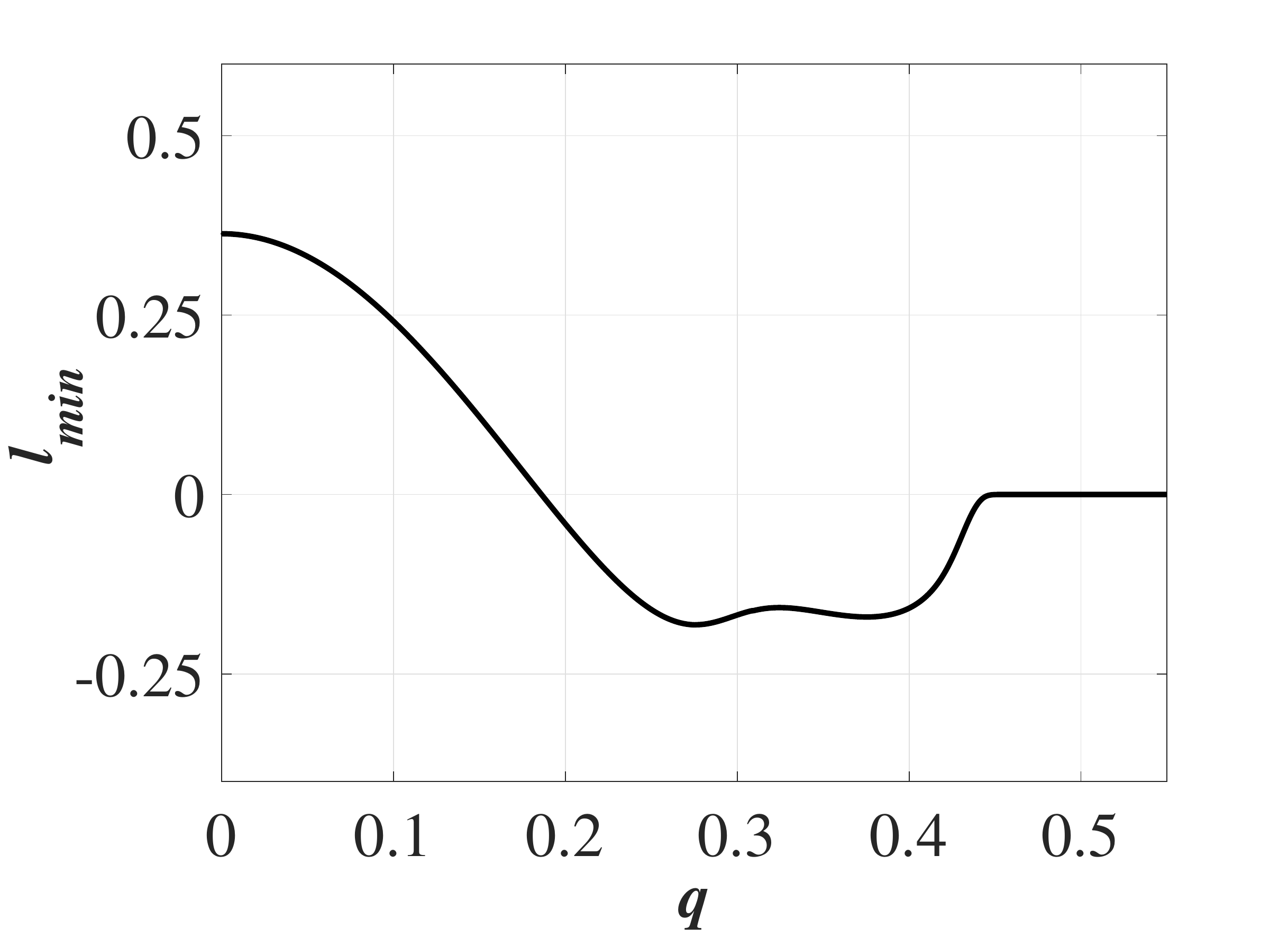}
\caption {The minimal eigenvalue $\l_{min}$ of the Hessian matrix as a function of $q$ for $\Gamma=0.07T_{c0}$ (left) and for $\Gamma=0.09T_{c0}$ (right) which correspond to a non-BTRS state with $s_{\pm}$ and $s_{++}$ symmetry, respectively.}
\label{stability_nonBTRS}
\end{figure}

As in the case of the BTRS state we consider the minimal eigenvalue $l_{min}$ of the Hessian matrix and plot it as a function on $q$ for the given value of $\Gamma=0.07T_{c0}$ and $\Gamma=0.09T_{c0}$ corresponding to $s_{\pm}$ and $s_{++}$ pairing symmetries, respectively (Fig. \ref{stability_nonBTRS}).

\subsection{The multiple-$q$ state}
A more complicated form of the Hessian matrix takes place for multiple-$q$ state because the GL free energy Eq.\ (\ref{GL_q1-q2}) is considered as a function of four variables $\Delta_1$, $\Delta_2$, $q_1$ and $q_2$. The calculation of the second derivatives yields
\begin{equation}
\label{Hessian_inhomo}
{H_{\left| {{\Delta _1}} \right|\left| {{\Delta _1}} \right|{q_1}{q_2}}} = \left( {\begin{array}{*{20}{c}}
  {\frac{{{\partial ^2}F}}{{{{\left| {{\Delta _1}} \right|}^2}}}}&{\frac{{{\partial ^2}F}}{{\left| {{\Delta _1}} \right|\left| {{\Delta _2}} \right|}}}&{\frac{{{\partial ^2}F}}{{\left| {{\Delta _1}} \right|{q_1}}}}&{\frac{{{\partial ^2}F}}{{\left| {{\Delta _1}} \right|{q_2}}}} \\ 
  {\frac{{{\partial ^2}F}}{{\left| {{\Delta _2}} \right|\left| {{\Delta _1}} \right|}}}&{\frac{{{\partial ^2}F}}{{{{\left| {{\Delta _2}} \right|}^2}}}}&{\frac{{{\partial ^2}F}}{{\left| {{\Delta _2}} \right|{q_1}}}}&{\frac{{{\partial ^2}F}}{{\left| {{\Delta _2}} \right|{q_2}}}} \\ 
  {\frac{{{\partial ^2}F}}{{{q_1}\left| {{\Delta _1}} \right|}}}&{\frac{{{\partial ^2}F}}{{{q_1}\left| {{\Delta _2}} \right|}}}&{\frac{{{\partial ^2}F}}{{q_1^2}}}&{\frac{{{\partial ^2}F}}{{{q_1}{q_2}}}} \\ 
  {\frac{{{\partial ^2}F}}{{{q_2}\left| {{\Delta _1}} \right|}}}&{\frac{{{\partial ^2}F}}{{{q_2}\left| {{\Delta _2}} \right|}}}&{\frac{{{\partial ^2}F}}{{{q_2}{q_1}}}}&{\frac{{{\partial ^2}F}}{{q_2^2}}} 
\end{array}} \right),
\end{equation}
where 
\begin{equation}
\frac{1}{L}\frac{{{\partial ^2}F}}{{{{\left| {{\Delta _1}} \right|}^2}}} = 2{a_{11}} + 6{b_{11}}{\left| {{\Delta _1}} \right|^2} + 2{b_{12}}{\left| {{\Delta _2}} \right|^2} + {k_{11}}{\hbar ^2}q_1^2 + 12{c_{11}}\left| {{\Delta _1}} \right|\left| {{\Delta _2}} \right|\frac{{\sin \left( {\left( {{q_1} - {q_2}} \right)L} \right)}}{{\left( {{q_1} - {q_2}} \right)L}} + {c_{12}}{\left| {{\Delta _2}} \right|^2}\frac{{\sin \left( {2\left( {{q_1} - {q_2}} \right)L} \right)}}{{\left( {{q_1} - {q_2}} \right)L}},
\end{equation}
\begin{equation}
\begin{gathered}
  \frac{1}{L}\frac{{{\partial ^2}F}}{{\partial \left| {{\Delta _1}} \right|\left| {{\Delta _2}} \right|}} = 4{b_{12}}\left| {{\Delta _1}} \right|\left| {{\Delta _2}} \right| + {k_{12}}{\hbar ^2}{q_1}{q_2}\frac{{\sin \left( {\left( {{q_1} - {q_2}} \right)L} \right)}}{{\left( {{q_1} - {q_2}} \right)L}} \hfill \\
   + 2\left( {{a_{12}} + 3{c_{11}}{{\left| {{\Delta _1}} \right|}^2} + 3{c_{22}}{{\left| {{\Delta _2}} \right|}^2}} \right)\frac{{\sin \left( {\left( {{q_1} - {q_2}} \right)L} \right)}}{{\left( {{q_1} - {q_2}} \right)L}} + 2{c_{12}}\left| {{\Delta _1}} \right|\left| {{\Delta _2}} \right|\frac{{\sin \left( {2\left( {{q_1} - {q_2}} \right)L} \right)}}{{\left( {{q_1} - {q_2}} \right)L}}, \hfill \\ 
\end{gathered}
\end{equation}
\begin{equation}
\begin{gathered}
  \frac{1}{L}\frac{{{\partial ^2}F}}{{\left| {{\Delta _1}} \right|{q_1}}} = 2{\mkern 1mu} {k_{11{\mkern 1mu} }}{\hbar ^2}\left| {{\Delta _1}} \right|{q_1} + \frac{{{k_{12{\mkern 1mu} }}{\hbar ^2}\left| {{\Delta _2}} \right|{\mkern 1mu} {\mkern 1mu} {q_2}{\mkern 1mu} \sin \left( {\left( {{q_1} - {q_2}} \right)L} \right)}}{{\left( {{q_1} - {q_2}} \right)L}} + \frac{{2{\mkern 1mu} {c_{12}}\left| {{\Delta _1}} \right|{\mkern 1mu} {{\left| {{\Delta _2}} \right|}^2}{\mkern 1mu} }}{{{q_1} - {q_2}}}\left( {\cos \left( {2{\mkern 1mu} \left( {{q_1} - {q_2}} \right)L} \right) - \frac{{\sin \left( {2\left( {{q_1} - {q_2}} \right)L} \right)}}{{2\left( {{q_1} - {q_2}} \right)L}}} \right) \hfill \\
   + \frac{{{k_{12}}{\hbar ^2}{\mkern 1mu} \left| {{\Delta _2}} \right|{\mkern 1mu} {q_1}{\mkern 1mu} {q_2} + 2{\mkern 1mu} \left( {{a_{12}}\left| {{\Delta _2}} \right| + 3{c_{11}}{\mkern 1mu} {{\left| {{\Delta _1}} \right|}^2}\left| {{\Delta _2}} \right|{\mkern 1mu}  + {c_{22}}{{\left| {{\Delta _2}} \right|}^3}{\mkern 1mu} } \right)}}{{{q_1} - {q_2}}}\left( {\cos\left( {\left( {{q_1} - {q_2}} \right)L} \right) - \frac{{\sin \left( {\left( {{q_1} - {q_2}} \right)L} \right)}}{{\left( {{q_1} - {q_2}} \right)L}}} \right), \hfill \\ 
\end{gathered}
\end{equation}
\begin{equation}
\begin{gathered}
  \frac{1}{L}\frac{{{\partial ^2}F}}{{\left| {{\Delta _1}} \right|{q_2}}} = \frac{{{k_{12{\mkern 1mu} }}{\hbar ^2}\left| {{\Delta _2}} \right|{\mkern 1mu} {\mkern 1mu} {q_1}{\mkern 1mu} \sin \left( {\left( {{q_1} - {q_2}} \right)L} \right)}}{{\left( {{q_1} - {q_2}} \right)L}} - \frac{{2{\mkern 1mu} {c_{12}}\left| {{\Delta _1}} \right|{\mkern 1mu} {{\left| {{\Delta _2}} \right|}^2}{\mkern 1mu} }}{{{q_1} - {q_2}}}\left( {\cos \left( {2{\mkern 1mu} \left( {{q_1} - {q_2}} \right)L} \right) - \frac{{{\mkern 1mu} \sin \left( {2\left( {{q_1} - {q_2}} \right)L} \right)}}{{2\left( {{q_1} - {q_2}} \right)L}}} \right) \hfill \\
   - \frac{{{k_{12}}{\hbar ^2}{\mkern 1mu} \left| {{\Delta _2}} \right|{\mkern 1mu} {q_1}{\mkern 1mu} {q_2} + 2{\mkern 1mu} \left( {{a_{12}}\left| {{\Delta _2}} \right| + 3{c_{11}}{\mkern 1mu} {{\left| {{\Delta _1}} \right|}^2}\left| {{\Delta _2}} \right|{\mkern 1mu}  + {c_{22}}{{\left| {{\Delta _2}} \right|}^3}{\mkern 1mu} } \right){\mkern 1mu} }}{{{q_1} - {q_2}}}\left( {\cos \left( {\left( {{q_1} - {q_2}} \right)L} \right) - \frac{{{\mkern 1mu} \sin \left( {\left( {{q_1} - {q_2}} \right)L} \right)}}{{\left( {{q_1} - {q_2}} \right)L}}} \right), \hfill \\ 
\end{gathered}
\end{equation}
\begin{equation}
\frac{1}{L}\frac{{{\partial ^2}F}}{{\partial {{\left| {{\Delta _2}} \right|}^2}}} = 2{a_{22}} + 6{b_{22}}{\left| {{\Delta _2}} \right|^2} + 2{b_{12}}{\left| {{\Delta _1}} \right|^2} + {k_{22}}{\hbar ^2}q_2^2 + 12{c_{22}}\left| {{\Delta _1}} \right|\left| {{\Delta _2}} \right|\frac{{\sin \left( {\left( {{q_1} - {q_2}} \right)L} \right)}}{{\left( {{q_1} - {q_2}} \right)L}} + {c_{12}}{\left| {{\Delta _1}} \right|^2}\frac{{\sin \left( {2\left( {{q_1} - {q_2}} \right)L} \right)}}{{\left( {{q_1} - {q_2}} \right)L}},
\end{equation}
\begin{equation}
\begin{gathered}
  \frac{1}{L}\frac{{{\partial ^2}F}}{{\left| {{\Delta _2}} \right|{q_1}}} = \frac{{{k_{12{\mkern 1mu} }}{\hbar ^2}\left| {{\Delta _1}} \right|{\mkern 1mu} {\mkern 1mu} {q_2}{\mkern 1mu} \sin \left( {\left( {{q_1} - {q_2}} \right)L} \right)}}{{\left( {{q_1} - {q_2}} \right)L}} + \frac{{2{\mkern 1mu} {c_{12}}{{\left| {{\Delta _1}} \right|}^2}{\mkern 1mu} \left| {{\Delta _2}} \right|{\mkern 1mu} }}{{{q_1} - {q_2}}}\left( {\cos \left( {2{\mkern 1mu} \left( {{q_1} - {q_2}} \right)L} \right) - \frac{{{\mkern 1mu} \sin \left( {2\left( {{q_1} - {q_2}} \right)L} \right)}}{{2\left( {{q_1} - {q_2}} \right)L}}} \right) \hfill \\
   + \frac{{{k_{12}}{\hbar ^2}{\mkern 1mu} \left| {{\Delta _1}} \right|{\mkern 1mu} {q_1}{\mkern 1mu} {q_2} + 2{\mkern 1mu} \left( {{a_{12}}\left| {{\Delta _1}} \right| + 3{c_{22}}{\mkern 1mu} \left| {{\Delta _1}} \right|{{\left| {{\Delta _2}} \right|}^2} + {c_{11}}{{\left| {{\Delta _1}} \right|}^3}{\mkern 1mu} } \right){\mkern 1mu} }}{{{q_1} - {q_2}}}\left( {\cos \left( {\left( {{q_1} - {q_2}} \right)L} \right) - \frac{{{\mkern 1mu} \sin \left( {\left( {{q_1} - {q_2}} \right)L} \right)}}{{\left( {{q_1} - {q_2}} \right)L}}} \right), \hfill \\ 
\end{gathered}
\end{equation}
\begin{equation}
\begin{gathered}
  \frac{1}{L}\frac{{{\partial ^2}F}}{{\left| {{\Delta _2}} \right|{q_2}}} = 2{\mkern 1mu} {k_{22{\mkern 1mu} }}{\hbar ^2}\left| {{\Delta _2}} \right|{q_2} + \frac{{{k_{12{\mkern 1mu} }}{\hbar ^2}\left| {{\Delta _1}} \right|{\mkern 1mu} {\mkern 1mu} {q_1}{\mkern 1mu} \sin \left( {\left( {{q_1} - {q_2}} \right)L} \right)}}{{\left( {{q_1} - {q_2}} \right)L}} - \frac{{2{\mkern 1mu} {c_{12}}{{\left| {{\Delta _1}} \right|}^2}{\mkern 1mu} \left| {{\Delta _2}} \right|{\mkern 1mu} }}{{{q_1} - {q_2}}}\left( {\cos \left( {2{\mkern 1mu} \left( {{q_1} - {q_2}} \right)L} \right) - \frac{{\sin \left( {2\left( {{q_1} - {q_2}} \right)L} \right)}}{{2\left( {{q_1} - {q_2}} \right)L}}} \right) \hfill \\
   - \frac{{{k_{12}}{\hbar ^2}{\mkern 1mu} \left| {{\Delta _1}} \right|{\mkern 1mu} {q_1}{\mkern 1mu} {q_2} + 2{\mkern 1mu} \left( {{a_{12}}\left| {{\Delta _1}} \right| + 3{c_{22}}{\mkern 1mu} \left| {{\Delta _1}} \right|{{\left| {{\Delta _2}} \right|}^2}{\mkern 1mu}  + {c_{11}}{{\left| {{\Delta _1}} \right|}^3}{\mkern 1mu} } \right)}}{{{q_1} - {q_2}}}\left( {\cos\left( {\left( {{q_1} - {q_2}} \right)L} \right) - \frac{{\sin \left( {\left( {{q_1} - {q_2}} \right)L} \right)}}{{\left( {{q_1} - {q_2}} \right)L}}} \right), \hfill \\ 
\end{gathered}
\end{equation}
\begin{equation}
\begin{gathered}
  \frac{1}{L}\frac{{{\partial ^2}F}}{{q_1^2}} = {k_{11}}{\hbar ^2}{\mkern 1mu} {\left| {{\Delta _1}} \right|^2} + \frac{{2{k_{12}}{\hbar ^2}\left| {{\Delta _1}} \right|\left| {{\Delta _2}} \right|{q_2}{\mkern 1mu} }}{{{q_1} - {q_2}}}\left( {\cos \left( {\left( {{q_1} - {q_2}} \right)L} \right) - \frac{{{\mkern 1mu} \sin \left( {\left( {{q_1} - {q_2}} \right)L} \right)}}{{\left( {{q_1} - {q_2}} \right)L}}} \right) \hfill \\
   - \frac{{{k_{12}}{\hbar ^2}\left| {{\Delta _1}} \right|\left| {{\Delta _2}} \right|{q_1}{\mkern 1mu} {q_2} + 2\left( {{a_{12}}\left| {{\Delta _1}} \right|\left| {{\Delta _2}} \right| + {c_{11}}{{\left| {{\Delta _1}} \right|}^3}\left| {{\Delta _2}} \right|{\mkern 1mu}  + {c_{22}}\left| {{\Delta _1}} \right|\left| {{\Delta _2}} \right|{{\mkern 1mu} ^3}} \right)}}{{{q_1} - {q_2}}} \hfill \\
   \times \left( {L\sin \left( {\left( {{q_1} - {q_2}} \right)L} \right) + \frac{{2{\mkern 1mu} \cos \left( {\left( {{q_1} - {q_2}} \right)L} \right)}}{{{q_1} - {q_2}}} - \frac{{2\sin \left( {\left( {{q_1} - {q_2}} \right)L} \right)}}{{{{\left( {{q_1} - {q_2}} \right)}^2}L}}} \right) \hfill \\
   - {\mkern 1mu} \frac{{2{c_{12}}{\mkern 1mu} {{\left| {{\Delta _1}} \right|}^2}{{\left| {{\Delta _2}} \right|}^2}}}{{{q_1} - {q_2}}}\left( {L\sin \left( {2{\mkern 1mu} \left( {{q_1} - {q_2}} \right)L} \right) + \frac{{\cos \left( {2{\mkern 1mu} \left( {{q_1} - {q_2}} \right)L} \right)}}{{{q_1} - {q_2}}} - \frac{{\sin \left( {2{\mkern 1mu} \left( {{q_1} - {q_2}} \right)L} \right)}}{{2{{\left( {{q_1} - {q_2}} \right)}^2}L}}} \right), \hfill \\ 
\end{gathered}
\end{equation}
\begin{equation}
\begin{gathered}
  \frac{1}{L}\frac{{{\partial ^2}F}}{{{q_1}{q_2}}} = \frac{{{k_{12}}{\hbar ^2}{\mkern 1mu} \left| {{\Delta _1}} \right|\left| {{\Delta _2}} \right|{\mkern 1mu} \sin \left( {\left( {{q_1} - {q_2}} \right)L} \right)}}{{\left( {{q_1} - {q_2}} \right)L}} - {k_{12}}{\hbar ^2}{\mkern 1mu} \left| {{\Delta _1}} \right|\left| {{\Delta _2}} \right|\left( {\cos \left( {\left( {{q_1} - {q_2}} \right)L} \right) - \frac{{\sin \left( {\left( {{q_1} - {q_2}} \right)L} \right)}}{{\left( {{q_1} - {q_2}} \right)L}}} \right) \hfill \\
   + \frac{{{k_{12}}{\hbar ^2}{\mkern 1mu} \left| {{\Delta _1}} \right|\left| {{\Delta _2}} \right|{\mkern 1mu} {q_1}{\mkern 1mu} {q_2} + 2\left( {{a_{12}}\left| {{\Delta _1}} \right|\left| {{\Delta _2}} \right| + {c_{11}}{{\left| {{\Delta _1}} \right|}^3}\left| {{\Delta _2}} \right|{\mkern 1mu}  + {c_{22}}\left| {{\Delta _1}} \right|\left| {{\Delta _2}} \right|{{\mkern 1mu} ^3}} \right){\mkern 1mu} }}{{{q_1} - {q_2}}} \hfill \\
   \times \left( {L\sin \left( {\left( {{q_1} - {q_2}} \right)L} \right) + \frac{{2{\mkern 1mu} {\mkern 1mu} \cos \left( {\left( {{q_1} - {q_2}} \right)L} \right)}}{{{q_1} - {q_2}}} - \frac{{2{\mkern 1mu} \sin \left( {\left( {{q_1} - {q_2}} \right)L} \right)}}{{{{\left( {{q_1} - {q_2}} \right)}^2}L}}} \right) \hfill \\
   + {\mkern 1mu} \frac{{2{c_{12}}{\mkern 1mu} {{\left| {{\Delta _1}} \right|}^2}{{\left| {{\Delta _2}} \right|}^2}}}{{{q_1} - {q_2}}}\left( {L\sin \left( {2{\mkern 1mu} \left( {{q_1} - {q_2}} \right)L} \right) + {\mkern 1mu} \frac{{\cos \left( {2{\mkern 1mu} \left( {{q_1} - {q_2}} \right)L} \right)}}{{{q_1} - {q_2}}} - \frac{{\sin \left( {2{\mkern 1mu} \left( {{q_1} - {q_2}} \right)L} \right)}}{{2{{\left( {{q_1} - {q_2}} \right)}^2}L}}} \right), \hfill \\ 
\end{gathered}
\end{equation}
\begin{equation}
\begin{gathered}
  \frac{1}{L}\frac{{{\partial ^2}F}}{{q_2^2}} = {k_{22}}{\hbar ^2}{\mkern 1mu} {\left| {{\Delta _2}} \right|^2} - \frac{{2{k_{12}}{\hbar ^2}\left| {{\Delta _1}} \right|\left| {{\Delta _2}} \right|{q_1}{\mkern 1mu} }}{{{q_1} - {q_2}}}\left( {\cos \left( {\left( {{q_1} - {q_2}} \right)L} \right) - \frac{{{\mkern 1mu} \sin \left( {\left( {{q_1} - {q_2}} \right)L} \right)}}{{\left( {{q_1} - {q_2}} \right)L}}} \right) \hfill \\
   - \frac{{{k_{12}}{\hbar ^2}\left| {{\Delta _1}} \right|\left| {{\Delta _2}} \right|{q_1}{\mkern 1mu} {q_2} + 2\left( {{a_{12}}\left| {{\Delta _1}} \right|\left| {{\Delta _2}} \right| + {c_{11}}{{\left| {{\Delta _1}} \right|}^3}\left| {{\Delta _2}} \right|{\mkern 1mu}  + {c_{22}}\left| {{\Delta _1}} \right|\left| {{\Delta _2}} \right|{{\mkern 1mu} ^3}} \right)}}{{{q_1} - {q_2}}} \hfill \\
   \times \left( {L\sin \left( {\left( {{q_1} - {q_2}} \right)L} \right) + \frac{{2{\mkern 1mu} \cos \left( {\left( {{q_1} - {q_2}} \right)L} \right)}}{{{q_1} - {q_2}}} - \frac{{2\sin \left( {\left( {{q_1} - {q_2}} \right)L} \right)}}{{{{\left( {{q_1} - {q_2}} \right)}^2}L}}} \right) \hfill \\
   - {\mkern 1mu} \frac{{2{c_{12}}{\mkern 1mu} {{\left| {{\Delta _1}} \right|}^2}{{\left| {{\Delta _2}} \right|}^2}}}{{{q_1} - {q_2}}}\left( {L\sin \left( {2{\mkern 1mu} \left( {{q_1} - {q_2}} \right)L} \right) + \frac{{\cos \left( {2{\mkern 1mu} \left( {{q_1} - {q_2}} \right)L} \right)}}{{{q_1} - {q_2}}} - \frac{{\sin \left( {2{\mkern 1mu} \left( {{q_1} - {q_2}} \right)L} \right)}}{{2{{\left( {{q_1} - {q_2}} \right)}^2}L}}} \right). \hfill \\ 
\end{gathered}
\end{equation}

\begin{figure}[H]
\includegraphics[width=0.325\columnwidth]{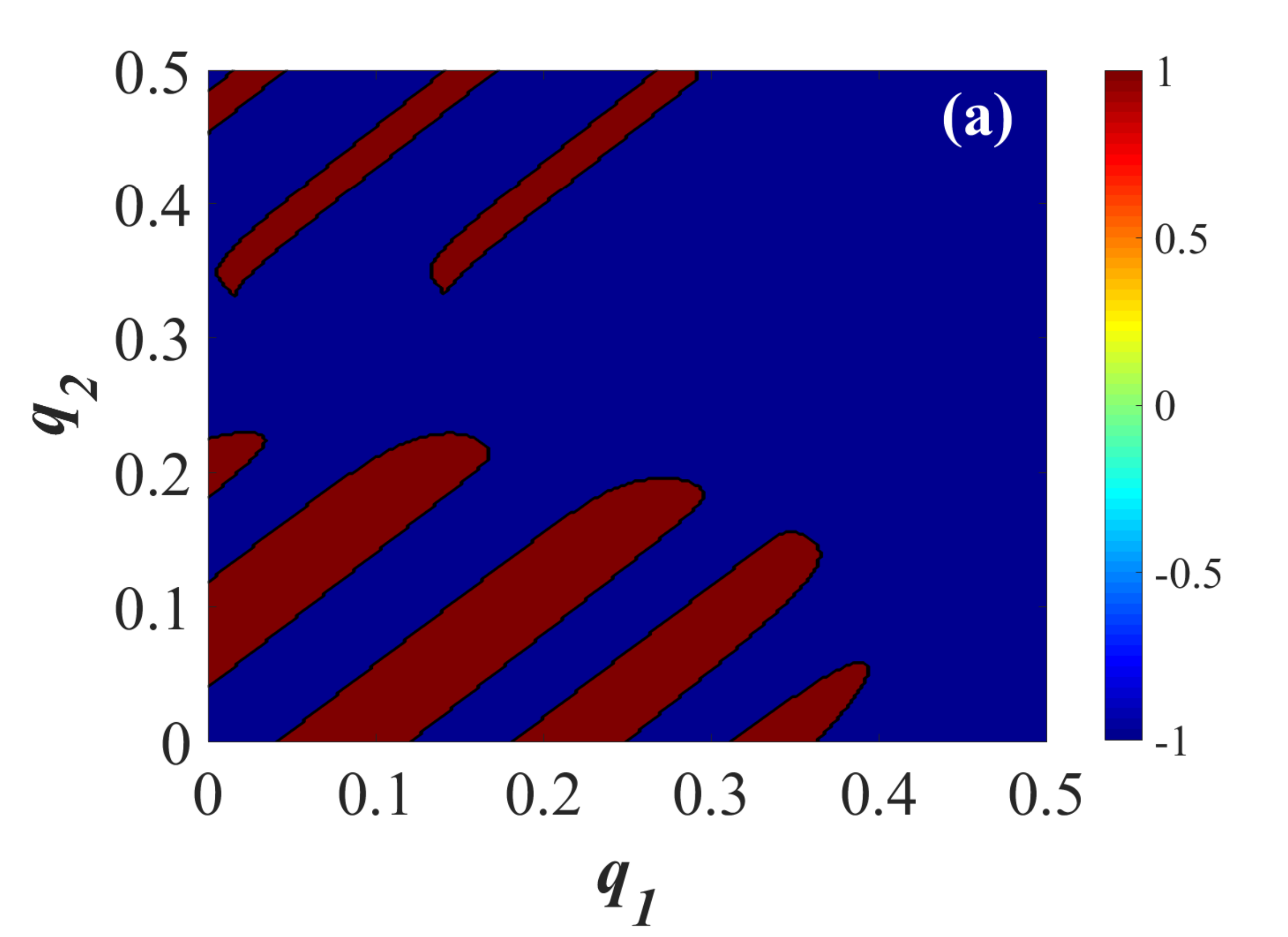}
\includegraphics[width=0.325\columnwidth]{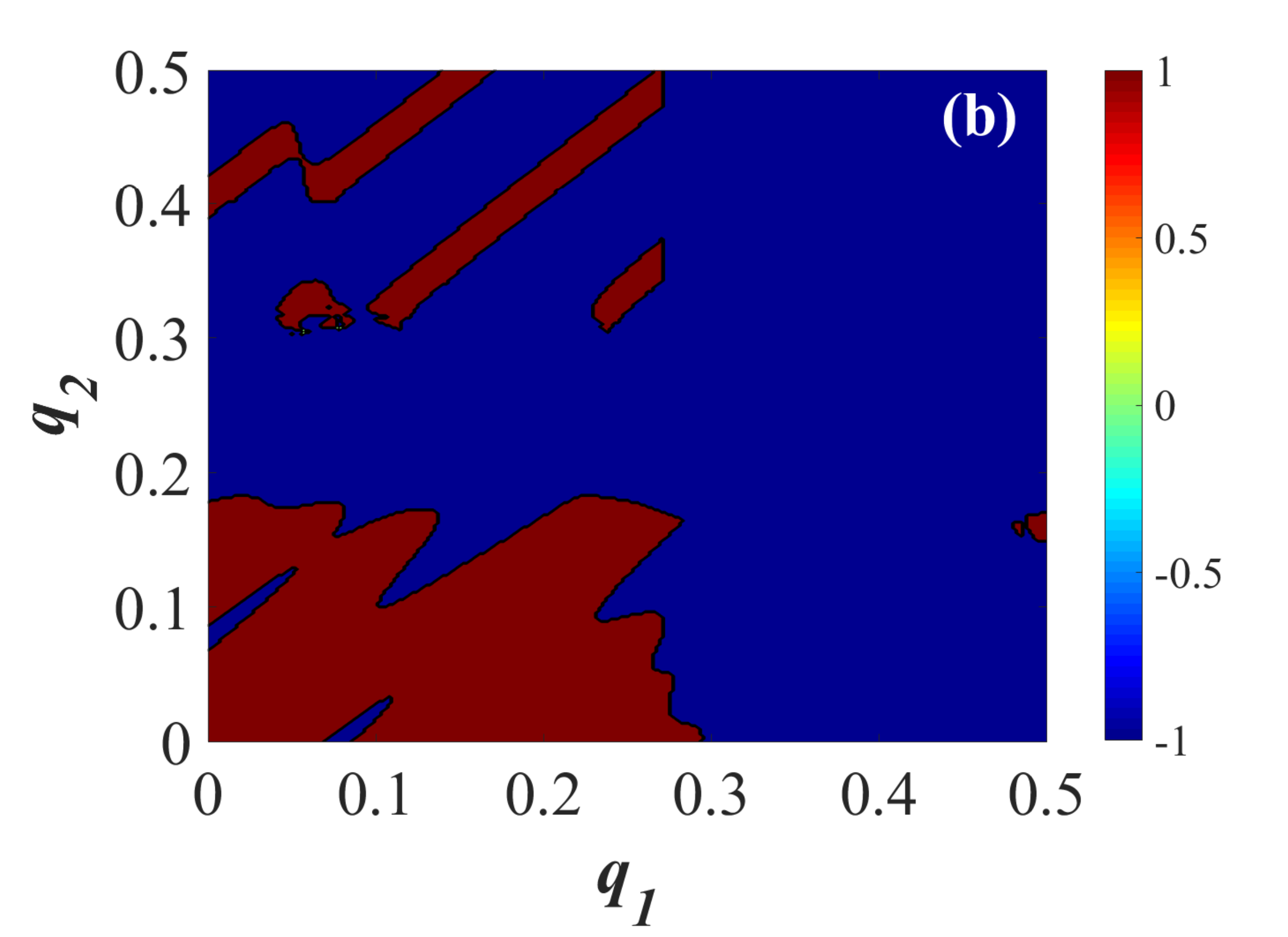}
\includegraphics[width=0.325\columnwidth]{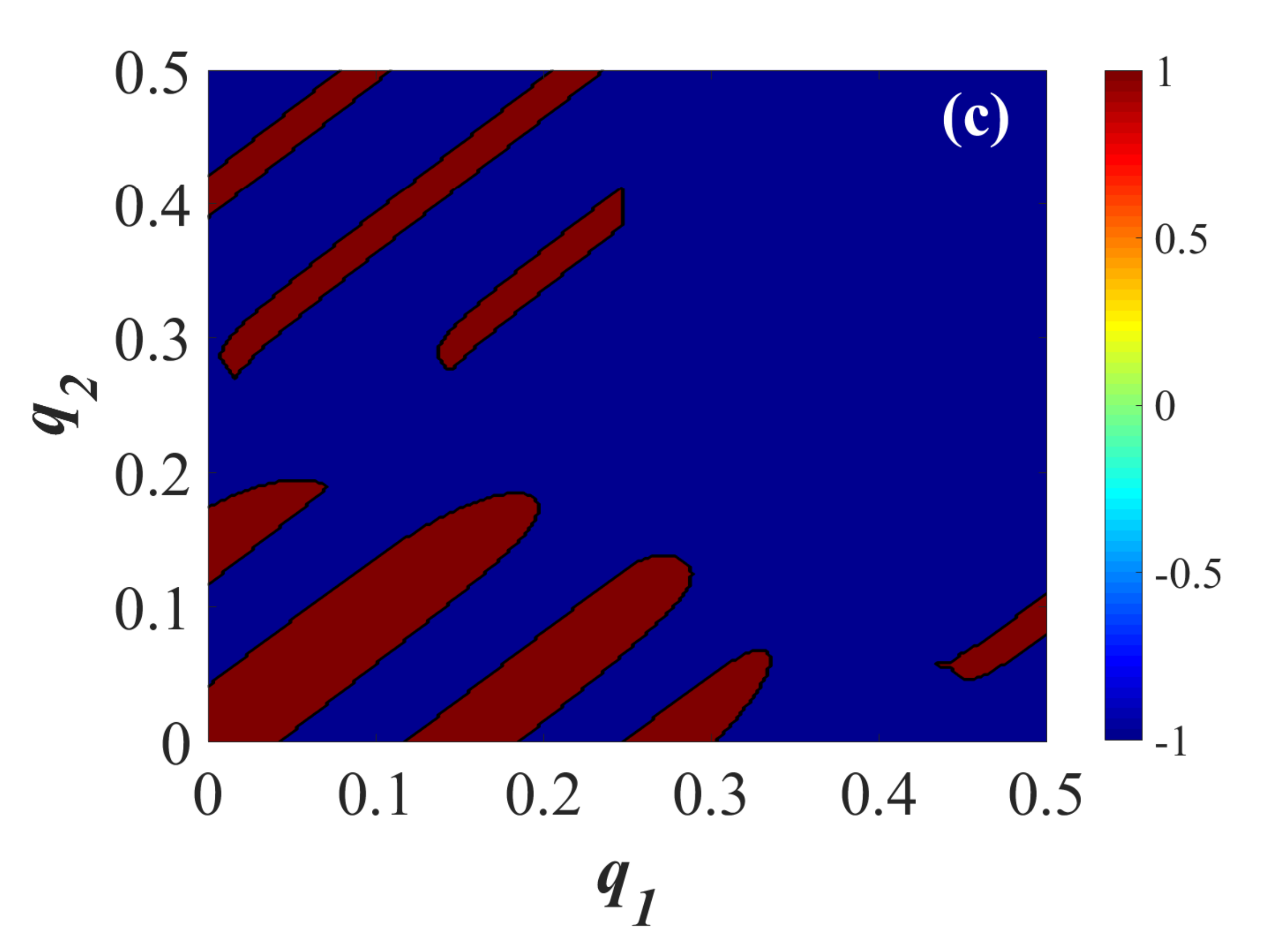}
\caption {The contour plot of the minimal eigenvalue $l_{min}$ of the Hessian matrix as a function of $q_1$ and $q_2$ for $\Gamma=0.07T_{c0}$ (a), $\Gamma=0.07982T_{c0}$ (b) and $\Gamma=0.09T_{c0}$ (c). Red regions signify the minimum of the GL free energy, while blue ones correspond to saddle-points or maxima.}
\label{stability_plots}
\end{figure}

Due to the presence of two superconducting momenta stability regions can be classified and analyzed in the form of the contour plot of the minimal eigenvalue $l_{min}$ as a function of $q_1$ and $q_2$ for a fixed value of $\Gamma$. The results for $\Gamma=0.07T_{c0}$, $\Gamma=0.07982T_{c0}$ and $\Gamma=0.09T_{c0}$ correspond to the selected earlier three reference points as presented in Figure \ref{stability_plots}).

\end{widetext}

\end{document}